\documentclass[prd,aps,nofootinbib,eqsecnum,superscriptaddress]{revtex4}
\usepackage[latin9]{inputenc}
\pdfoutput=1
\usepackage{amsmath}
\usepackage{amssymb}
\usepackage{graphicx}
\usepackage[unicode=true,
 bookmarks=false,
 breaklinks=false,pdfborder={0 0 1},backref=section,colorlinks=false]
 {hyperref}
\hypersetup{
 colorlinks}

\makeatletter
\@ifundefined{textcolor}{}
{%
 \definecolor{BLACK}{gray}{0}
 \definecolor{WHITE}{gray}{1}
 \definecolor{RED}{rgb}{1,0,0}
 \definecolor{GREEN}{rgb}{0,1,0}
 \definecolor{BLUE}{rgb}{0,0,1}
 \definecolor{CYAN}{cmyk}{1,0,0,0}
 \definecolor{MAGENTA}{cmyk}{0,1,0,0}
 \definecolor{YELLOW}{cmyk}{0,0,1,0}
}

\usepackage{bm}\usepackage[all]{hypcap}
\usepackage{psfrag}\usepackage{epsfig}\usepackage{mathrsfs}
\usepackage{wasysym}\usepackage{epstopdf}
\usepackage{comment}
\usepackage{pbox}
\usepackage{array}
\usepackage{xspace}
\usepackage[usenames,dvipsnames]{color}

\def\addUPitt{Pittsburgh Particle Physics Astrophysics and Cosmology Center, Department of Physics and Astronomy, University of Pittsburgh, Pittsburgh, PA 15260, USA}

\@ifundefined{showcaptionsetup}{}{%
 \PassOptionsToPackage{caption=false}{subfig}}
\usepackage{subfig}
\makeatother

\begin{document}
\title{Next-to-leading order spin-orbit effects in the equations of motion,
energy loss and phase evolution of binaries of compact bodies in the
effective field theory approach}
\author{Brian Pardo}
\affiliation{\addUPitt}
\author{Nat\'alia T.~Maia}
\affiliation{\addUPitt}
\begin{abstract}
We compute spin-orbit effects in the equations of motion, binding energy
and energy loss of binary systems of compact objects at the next-to-leading
order in the post-Newtonian (PN) approximation in the effective field
theory (EFT) framework. We then use these quantities to compute the
evolution of the orbital frequency and accumulated orbital phase including
spin-orbit effects beyond the dominant order. To obtain the results
presented in this paper, we make use of known ingredients in the EFT
literature, such as the potential and the multipole moments with spin
effects at next-to-leading order, and which are given in the linearized
harmonic gauge and with the spins in the locally flat frame. We also
obtain the correction to the center-of-mass frame caused by spin-orbit
effects at next-to-leading order. We demonstrate the equivalence between
our EFT results and those which were obtained elsewhere using different
formalisms. The results presented in this paper provide us with the
final ingredients for the construction of theoretical templates for
gravitational waves including next-to-leading order spin-orbit effects,
which will be presented in a future publication.
\end{abstract}
\maketitle

\section{Introduction}

Gravitational wave astronomy is based on high precision experimental
and theoretical physics. The main sources of gravitational wave signals
which can be detected by the ground-based observatories LIGO and Virgo
\cite{ligo1,ligo2,ligo3,ligo4,ligo5,ligo6,ligo7,ligo8,ligo9,ligo10,ligo11}
are binary systems of compact objects. During the inspiral stage,
those systems can be studied analytically through perturbative approaches
such as the post-Newtonian (PN) approximation, which uses the ratio
between the relative velocity and the speed of light $(v^{2}/c^{2})$
as the expansion parameter. For accuracy, the calculations have to
be carried out to high orders in the expansion parameter in order
to be valid up to the late inspiral stage, which is when the theoretical
templates are matched onto the numerical ones. In fact, interesting
physics can be studied only when we go beyond the leading order, for
instance finite size effects such as spin, which plays an important
role in understanding the formation and the evolution of binary systems
\cite{Spin_Offner_2016,Spin_Sedda_2018,Spin_Sedda_2020}.

The EFT framework we use in this paper, called non-relativistic general
relativity (NRGR), originally proposed in \cite{nrgr} and extended
to accommodate rotating objects in \cite{nrgrs}, is an independent
approach to the investigation of the dynamics of binaries of compact
objects. The current state of the art for the EFT formalism is 4PN
order \cite{rec1_Galley_2016,rec2_Foffa_2019,rec3_Porto_2017,rec4_Foffa_2019}
in the conservative sector for non-spinning bodies. The spin sector
of this formalism--the focus of this paper--has also seen extensive
development in the past 15 years. The leading order (LO) spin effects
in the conservative dynamics were derived through the NRGR formalism
in \cite{nrgrs}, while the next-to-leading order (NLO) and next-to-next-to
leading order (N$^{2}$LO) spin effects were studied in \cite{nrgrso,nrgrss,nrgrs2,eih,Levi:2008nh,Levi:2010zu}
and \cite{levinnlo2,levinnlo3}, respectively. Recently, the next-to-next-to-next-to-leading
order (N$^{3}$LO) gravitational spin-orbit \cite{levi2020nnnloso}
and quadratic-in-spin \cite{levi2020nnnloss} interactions were also
investigated. Beyond the linear and the quadratic spin effcets, the
LO cubic and quartic spin interactions \cite{levi3} and the NLO cubic
spin interactions \cite{levi2019NLOs3} were also explored via the
NRGR framework. In the radiative sector of this formalism, spin effects
in the multipole moments were obtained in \cite{Porto:2010zg,Porto:2012as},
and the LO spin effects in radiation reaction were computed in \cite{paper1,paper2}.

Although crucial ingredients for the description of the dynamics of
binaries of compact bodies including NLO spin effects were previously
computed using the NRGR formalism, in particular the spin-orbit potential
and the spin evolution in \cite{nrgrso}, and the multipole moments
in \cite{Porto:2010zg}, other important quantities associated to
NLO spin-orbit effects--such as the equations of motion of the compact
bodies, the system's binding energy, its energy loss and phase evolution--have
yet to be derived in the NRGR framework. One of the purposes of this
paper is to obtain those quantities, since they play an important
role in the investigation of the physics of binary systems. For instance,
the acceleration we derive in this paper, which composes a 2.5PN correction
to the system's equations of motion, is used to compute the energy
loss associated to the emission of gravitational waves but also to
obtain the phase evolution of the binary system. In addition, this
acceleration is a necessary ingredient for the construction of theoretical
templates of gravitational waves accounting for NLO spin-orbit effects,
which shall be presented in a future publication. Another spin-orbit
effect that enters at 2.5PN order is the correction to the center-of-mass
frame. Although it does not affect the NLO spin-orbit acceleration
obtained here, we derive this 2.5PN spin-orbit correction to the center-of-mass
for completeness, with the intent to provide the final pieces related
to NLO spin-orbit effects in order to allow the EFT calculations to
continue without impediment at higher orders. Furthermore, we provide
a discussion between the results obtained in this paper and the ones
in the literature \cite{buo1,buo2}, where different gauge and spin
definitions are used while a more traditional PN approach to general
relativity is followed, and we show that, through a redefinition of
the spin variables, equivalence can be proven even before gauge invariant
quantities are computed. This present paper, therefore, also serves
as a demonstration of the equivalence between the NRGR methodology
and more traditional approaches to general relativity up to NLO regarding
spin-orbit effects, both in the conservative and dissipative sectors.

We organize this paper as follows. In section \ref{sec:EFTsetup},
we provide a brief summary of the NRGR formalism (we recommend \cite{nrgrLH,Rothstein:2014sra,Foffa:2013qca,Porto:2016pyg,Levi:2018nxp}
for a comprehensive review). We derive the NLO spin-orbit acceleration
in section \ref{sec:EOM} by computing the Euler-Lagrange equations
of the potential obtained in \cite{nrgrso} but also by extracting
contributions coming from order reducing terms in lower order accelerations
and from spin precession and constraints. In section \ref{sec:Energy},
we take the Legendre transform of the potential derived in \cite{nrgrso}
to obtain the NLO spin-orbit effects in the binding energy of the
binary system, and we make use of the acceleration computed in section
\ref{sec:EOM} as well as the multipole moments obtained in \cite{Porto:2010zg}
to calculate the energy loss due to the emission of gravitational
radiation. Then, in section \ref{sec:Phase} we use the results obtained
in the sections \ref{sec:EOM} and \ref{sec:Energy} to calculate
the evolution of the orbital frequency of the binary system and its
phase evolution accounting for NLO spin-orbit effects for quasi-circular
orbits within the adiabatic approximation. In section \ref{sec:CM}
we compute the NLO spin-orbit effects in the $00$-component of the
binary's pseudotensor, which we use to extract the NLO spin-orbit
correction to the center-of-mass frame by Taylor expanding its expression
up to the first order in the radiation field momentum. In section
\ref{sec:Comparison}, we discuss the specific spin transformations
which map the main results of this paper to those in the literature
obtained from traditional PN approaches. In section \ref{sec:Final-remarks},
we provide the reader with our final remarks on the contributions
of this paper. We compile the known ingredients used to derive the
results of this paper in appendix \ref{sec:App-A} for convenience.

A number of conventions and definitions are utilized throughout this
paper. The masses $m_{1}$ and $m_{2}$ of the binary components are
used to define the following quantities: $m\equiv m_{1}+m_{2}$, $\nu\equiv m_{1}m_{2}/m^{2}$,
and $\mu\equiv m\nu$. The relative position is defined as $\mathbf{r}\equiv\mathbf{x}_{1}-\mathbf{x}_{2}$
and its unit vector given by $\mathbf{n}\equiv\mathbf{r}/r$; thus
$\mathbf{v}\equiv\mathbf{v}_{1}-\mathbf{v}_{2}$ and $\mathbf{a}\equiv\mathbf{a}_{1}-\mathbf{a}_{2}$
are the relative velocity and acceleration, respectively. If those
relative quantities appear inside a sum over the compact objects indices
$A,B=1,2$, they should be considered as dependening on the those
indices instead, e.g. $\mathbf{r}=\mathbf{x}_{A}-\mathbf{x}_{B}$.
We use the Newtonian orbital angular momentum vector defined by $\mathbf{L}\equiv m\nu\mathbf{r}\times\mathbf{v}$.
We use the spins $\mathbf{S}_{1}$ and $\mathbf{S}_{2}$ of the bodies
to define
\begin{align}
\mathbf{S} & \equiv\mathbf{S}_{1}+\mathbf{S}_{2},\label{eq:S}\\
\boldsymbol{\mathbf{\Sigma}} & \equiv m\biggl(\frac{\mathbf{S}_{2}}{m_{2}}-\frac{\mathbf{S}_{1}}{m_{1}}\biggr),\label{eq:Sigma}
\end{align}
which are two useful quantities to write results in a more elegant
way. We adopt the mostly minus signature $\left(1,-1,-1,-1\right)$
for the Minkowskian metric $\eta^{\alpha\beta}$. We use $c=1$ units
and the Planck mass is defined as $m_{\text{Pl}}\equiv1/\sqrt{32\pi G}$.


\section{NRGR setup\label{sec:EFTsetup}}

\subsection{Conservative sector}

The EFT approach is well suited to investigate the inspiral stage
of the binary system, when there is a clear hierarchy between the
length scales of the system: the size of the compact objects $r_{s}$,
the orbital separation $r$, and the radiation wavelength. The modes
of the perturbation $h_{\mu\nu}$ of the gravitational field, $g_{\mu\nu}=\eta_{\mu\nu}+h_{\mu\nu}$,
can be split into two different components: $h_{\mu\nu}=H_{\mu\nu}+\bar{h}_{\mu\nu}$,
where $H_{\mu\nu}$ are off-shell \textit{potential} modes of the
field which mediate gravitational attraction, and $\bar{h}_{\mu\nu}$
represent the on-shell propagating \textit{radiation} modes generated
by the motion of the compact bodies in the binary \cite{nrgr}. Then,
we start with the full theory action
\begin{equation}
S=S_{\text{EH}}+S_{\text{gf}}+S_{\text{pp}}+S_{\textrm{sg}}+\cdots,\label{eq:full}
\end{equation}
where 
\begin{align}
S_{\text{EH}} & =-2m_{\text{Pl}}\int d^{4}x\,\sqrt{-g}\:g_{\mu\nu}R^{\mu\nu},\label{eq:EH}\\
S_{\text{gf}} & =\int d^{4}x\,\sqrt{-\bar{g}}\,\bar{\Gamma}^{\mu}\bar{\Gamma}_{\mu},\label{eq:gf}\\
S_{\text{pp}} & =-\sum_{A}m_{A}\int d\tau_{A},\label{eq:pp}\\
S_{\textrm{sg}} & =-\frac{1}{2}\sum_{A}\int dt\,v_{A}^{\mu}\omega_{\mu ab}S_{A}^{ab}.\label{eq:sg}
\end{align}
The Einstein-Hilbert action \eqref{eq:EH} represents the purely gravitational
interaction terms. We utilize the linearized harmonic gauge fixing
action \eqref{eq:gf}, which is given in terms of the background field
metric $\bar{g}_{\mu\nu}\equiv\eta_{\mu\nu}+\bar{h}_{\mu\nu}$, in
order to maintain the diffeomorphism invariance even after the potential
modes of the gravitational fields are integrated out. Thus, we have
$\bar{\Gamma}_{\mu}\equiv\bar{\nabla}_{\alpha}H_{\mu}^{\alpha}-\frac{1}{2}\bar{\nabla}_{\mu}H_{\alpha}^{\alpha},$
with $\bar{\nabla}_{\mu}$ representing the covariant derivative associated
to the background metric $\bar{g}_{\mu\nu}$. The point particle approximation
is used to describe the two constituents of the binary system, hence
\eqref{eq:pp}; the index $A=1,2$ is a label for the two compact
bodies.

The final term \eqref{eq:sg} represents the spin-gravity coupling.
Choosing the coordinate time $t$ as the worldline parameter, the
spin action is composed of the four-velocity of the compact bodies
$v_{A}^{\mu}$, the spin connection $\omega_{\mu ab}\equiv e_{\nu b}\nabla_{\mu}e_{a}^{\nu}$,\footnote{This definition differs by a minus sign from the standard convention.}
and the antisymmetric spin tensor $S_{A}^{ab}\equiv S_{A}^{\mu\nu}e_{\mu}^{a}e_{\nu}^{b}$
given in the locally flat frame. The vierbien $e_{\mu}^{a}$ is defined
such that $e_{\mu}^{a}e_{\nu}^{b}\eta_{ab}=g_{\mu\nu}$ and $\nabla_{\mu}$
is the covariant derivative associated with the metric $g_{\mu\nu}$.
The locally flat frame retains a residual Lorentz invariance, and
is equivalent to adding an additional element of the SO$(3,1)$ group
to the worldline to implement rotations \cite{nrgrs}. Finally, the
ellipsis in \eqref{eq:full} represents other interactions that we
are not accounting for in this present paper, including finite size
terms which are quadratic or higher in the spins.

The Feynman rules for this EFT theory are obtained after imposing
the low velocity limit and the weak field approximation in the full
action \eqref{eq:full}. The derivatives of the off-shell potential
modes scale as $\partial_{0}H_{\mu\nu}\sim\left(\frac{v}{r}\right)H_{\mu\nu}$
and $\partial_{i}H_{\mu\nu}\sim\left(\frac{1}{r}\right)H_{\mu\nu}$
while derivatives of the on-shell radiation modes scale as $\partial_{\alpha}\bar{h}_{\mu\nu}\sim\left(\frac{v}{r}\right)\bar{h}_{\mu\nu}$.
For maximally rotating objects, we power count the spin as $S\sim Lv$,
where $L$ is the angular momentum. Therefore, we can determine how
each term in the full action scales with respect to the expansion
parameter $v^{2}$. This power counting allows us to systematically
compute spin or other effects at any desired order in the PN expansion.

After imposing the weak-field approximation using
\begin{equation}
e_{\mu}^{a}=\delta_{\mu}^{a}+\frac{1}{2}\delta_{\nu}^{a}\biggl(h_{\enskip\mu}^{\nu}-\frac{1}{4}h_{\enskip\rho}^{\nu}h_{\enskip\mu}^{\rho}\biggr)+\cdots,\label{eq:e_expanded}
\end{equation}
the spin-gravity Lagrangian becomes an infinite series of terms with
a single spin tensor contracted with the gravitational field at different
orders in its perturbation:
\begin{equation}
\ensuremath{L_{\textrm{sg}}=\sum_{A=1,2}\biggl[\frac{1}{2m_{\mathrm{Pl}}}\delta_{a}^{\alpha}\delta_{b}^{\beta}h_{\alpha\gamma,\beta}v_{A}^{\gamma}S_{A}^{ab}+\frac{1}{4m_{\mathrm{Pl}}^{2}}\delta_{a}^{\beta}\delta_{b}^{\gamma}h_{\enskip\gamma}^{\lambda}\biggl(\frac{1}{2}h_{\beta\lambda,\mu}+h_{\mu\lambda,\beta}-h_{\mu\beta,\lambda}\biggr)v_{A}^{\mu}S_{A}^{ab}+\cdots\biggr]}.\label{eq:Lsg}
\end{equation}
From this Lagrangian, we can extract all the relevant couplings that
are needed at the PN order that we consider in this paper. Moreover,
when we split the weak field into the two different modes, we can
obtain the potential--from which the spin-orbit equations of motion
can be derived--by integrating out the potential modes of the gravitational
field. Both the potential from \cite{nrgrso} and the couplings needed
to compute NLO spin-orbit effects, which are 2.5PN corrections in
the equations of motion, the binding energy and the center-of-mass
position, are presented in appendix~\ref{sec:App-A}.

Using a rank-2 antisymmetric tensor to describe spin in a four dimensional
spacetime comes with a cost: there are a total of six independent
degrees of freedom to play the role of the three necessary angles
to describe the rotation of a body. For the purpose of eliminating
the three unphysical components of the spin tensor, we impose constraints
known as spin supplementary conditions (SSC). In this paper, we use
the covariant SSC, which is given by the contraction of the spin tensor
with the linear momentum
\begin{equation}
p_{a}S^{ab}=0.\label{eq:cov}
\end{equation}
Even though the bodies label has been suppressed in the equation above,
notice that this constraint must be imposed for each of the compact
bodies.

\subsection{Radiative sector\label{subsec:Radiative-sector}}
The long-wavelength effective theory can be constructed by integrating
out the potential modes of the gravitational field. The binary system
is then described as a single point-like object endowed with a series
of multipole moments \cite{andirad,Porto:2016pyg}:
\begin{equation}
S_{\text{eff}}^{\text{rad}}\left[\bar{h},\mathbf{x}_{a}\right]=\int dt\,\sqrt{\bar{g}_{00}}\,\biggl[-M(t)+\sum_{l=2}^{\infty}\biggl(\frac{1}{l!}I^{L}\nabla_{L-2}E_{i_{l-1}i_{l}}-\frac{2l}{(2l+1)!}J^{L}\nabla_{L-2}B_{i_{l-1}i_{l}}\biggr)\biggr].\label{eq:Srad}
\end{equation}
The center of mass of the binary system is placed at the origin and
at rest with respect to distant observers, such that $dt\sqrt{\bar{g}_{00}}=d\tau$,
while $M(t)$ is the Bondi mass of the binary system. In the action
above, the electric and the magnetic components of the Weyl tensor
are coupled to the mass and current multipole moments, respectively.
Notice that a multi-index representation $L=i_{1}\ldots i_{l}$ is
used. The general expressions of the multipole moments $I^{L}$ and
$J^{L}$ in terms of the components of the pseudotensor of the binary
system, which can be found in \cite{andirad2}, are determined by
matching the effective action \eqref{eq:Srad} in the long wavelength
limit onto the full action valid below the orbital scale \eqref{eq:full}.
On the other hand, the pseudotensor $T^{\mu\nu}$, which satisfies
the conservation law $\partial_{\mu}T^{\mu\nu}=0$, can be read off
from
\begin{equation}
\Gamma\left[\bar{h}\right]=-\frac{1}{2m_{\text{Pl}}}\int d^{4}x\,T^{\mu\nu}\bar{h}_{\mu\nu},\label{eq:Sonegrav-1-1}
\end{equation}
when we integrate out the potential modes in the full action \eqref{eq:full}
for all terms containing a single radiation field.

The knowledge of the components of the pseudotensor and, consequently,
of the multipole moments of the binary system, is required in order
to determine the energy which is lost in the emission of gravitional
waves \cite{andirad2}:
\begin{equation}
\frac{dE}{dt}=-\frac{G}{5}\left(I_{ij}^{(3)}I_{ij}^{(3)}+\frac{16}{9}J_{ij}^{(3)}J_{ij}^{(3)}+\frac{5}{189}I_{ijk}^{(4)}I_{ijk}^{(4)}+\frac{5}{84}J_{ijk}^{(4)}J_{ijk}^{(4)}+\cdots\right).\label{eq:dEdt}
\end{equation}
All the necessary multipole moments for the computation of the NLO
spin-orbit effects in the energy loss, which we compute in the section
\ref{subsec:Energy-Loss} of this paper, were computed in \cite{Porto:2010zg}
and are presented in the center-of-mass frame in appendix \ref{sec:App-A}.

\section{Equations of motion\label{sec:EOM}}

In the PN approximation, the acceleration of the constituents of the
binary systems is given as a series of relativistic corrections to
the dominant Newtonian gravitational acceleration. If we disregard,
for the purposes of this paper, radiation reaction and effects of
quadratic (or higher) order in the spins, the acceleration can be
presented as\footnote{The first quadratic spin effects enter at 2PN order, while radiation
reaction enters at 2.5PN order.}:
\begin{equation}
\mathbf{a}=\mathbf{a}^{\text{(0PN)}}+\mathbf{a}^{\text{(1PN)}}+\mathbf{a}_{\text{SO}}^{\text{(1.5PN)}}+\mathbf{a}^{\text{(2PN)}}+\mathbf{a}_{\text{SO}}^{\text{(2.5PN)}}+\cdots.\label{eq:A_PNexpanded}
\end{equation}
The expressions for the non-spin accelerations in the right hand side
of the equation above are given in appendix \ref{sec:App-A}. The
LO spin-orbit acceleration--a 1.5PN correction to the equation of
motion--can be derived from the potential $V_{\textrm{1.5PN}}^{\textrm{SO}}$
given in \eqref{eq:15PNpot}. Computing the Euler-Lagrange equations
using that potential gives
\begin{align}
(\mathbf{a}_{1}^{i})^{V_{\textrm{SO}}^{\textrm{(1.5PN)}}} & =\frac{G}{r^{3}}\biggl\{\frac{m_{2}}{m_{1}}\left[(3\mathbf{n}\dot{r}-2\mathbf{v}_{1}+3\mathbf{v}_{2})\times\mathbf{S}_{1}\right]^{i}+\left[(6\mathbf{n}\dot{r}-4\mathbf{v}_{1}+3\mathbf{v}_{2})\times\mathbf{S}_{2}\right]^{i}\nonumber \\
 & \qquad\qquad-\biggl[\mathbf{r}\times\biggl(\frac{m_{2}}{m_{1}}\dot{\mathbf{S}}_{1}+2\dot{\mathbf{S}}_{2}\biggr)\biggr]^{i}+\biggl[S_{2}^{i0}-\frac{m_{2}}{m_{1}}S_{1}^{i0}+3\mathbf{n}^{i}\mathbf{n}^{j}\biggl(\frac{m_{2}}{m_{1}}S_{1}^{j0}-S_{2}^{j0}\biggr)\biggr]_{\text{cov}}\nonumber \\
 & \qquad\qquad+3\mathbf{n}^{i}\mathbf{n}\cdot\biggl(\frac{m_{2}}{m_{1}}\mathbf{v}_{1}\times\mathbf{S}_{1}-2\frac{m_{2}}{m_{1}}\mathbf{v}_{2}\times\mathbf{S}_{1}+2\mathbf{v}_{1}\times\mathbf{S}_{2}-\mathbf{v}_{2}\times\mathbf{S}_{2}\biggr)\biggr\}.\label{eq:a1(1.5PN)gen}
\end{align}

Notice that the expression above is given in a general form: it includes
time derivatives of the spin vectors, which actually contribute only
at orders higher than 1.5PN since $\smash{\dot{S}\sim\frac{v^{3}}{r}S}$;
it also shows the explicit dependence on the $\smash{S_{1,2}^{j0}}$
variables, which will be removed by enforcing the covariant SSC \eqref{eq:cov}.
Although we kept $\smash{S_{1,2}^{j0}}$ variables to indicate that
those terms will also contribute to orders higher than 1.5PN due to
PN corrections in the covariant SSC, the result in \eqref{eq:a1(1.5PN)gen}
is \emph{only} valid in the covariant SSC and is not general to other
choices of constraints\footnote{If we were working with the Newton-Wigner SSC, for instance, we would
have to impose the constraint at the level of the potential before
computing the Euler-Lagrange equations. See the discussion presented
in appendix E of \cite{nrgrs} for more details.} . Up to 1PN order, the spin tensors can be written in terms of the
spin vectors in the covariant SSC as
\begin{equation}
S_{A}^{0i}=\mathbf{S}_{A}\times\mathbf{v}_{A}+\frac{2Gm_{B}}{r}\mathbf{S}_{A}\times\mathbf{v}+\mathcal{O}(\mathbf{S}^{2})\label{eq:CovSi0}
\end{equation}
and
\begin{equation}
S_{A}^{ij}=\epsilon^{ijk}\mathbf{S}_{A}^{k}.\label{eq:CovSij}
\end{equation}
Therefore, after imposing the covariant SSC in \eqref{eq:a1(1.5PN)gen}
and keeping only terms which enter at the lowest PN order, we can
write the well-defined expression for the LO spin-orbit acceleration
\cite{nrgrs}:
\begin{align}
(\mathbf{a}_{1}^{i})_{\text{SO}}^{\text{(1.5PN)}} & =\frac{G}{r^{3}}\biggl\{3\frac{m_{2}}{m_{1}}[(\mathbf{S}_{1}\times\mathbf{v})^{i}-\dot{r}(\mathbf{S}_{1}\times\mathbf{n})^{i}-2\mathbf{S}_{1}\cdot(\mathbf{v}\times\mathbf{n})\mathbf{n}^{i}]\nonumber \\
 & \qquad\qquad+4(\mathbf{S}_{2}\times\mathbf{v})^{i}-6\dot{r}(\mathbf{S}_{2}\times\mathbf{n})^{i}-6\mathbf{S}_{2}\cdot(\mathbf{v}\times\mathbf{n})\mathbf{n}^{i}\biggr\}.\label{eq:a1(1.5PN)-1}
\end{align}

The purpose of this section is to advance to the next step, namely,
to obtain the equations of motion linear in the spins for the binary
system at 1PN beyond equation \eqref{eq:a1(1.5PN)-1}, which is a
2.5PN correction to the Newtonian acceleration. The result for the
NLO spin-orbit acceleration can be presented as the sum of two distinct
contributions:
\begin{equation}
(\mathbf{a}_{1}^{i})_{\text{SO}}^{\text{(2.5PN)}}=(\mathbf{a}_{1}^{i})^{V_{\textrm{SO}}^{\textrm{(2.5PN)}}}+(\mathbf{a}_{1}^{i})^{\text{(Red.)}}.\label{eq:A1(2.5PN)gen}
\end{equation}
The first term in the right hand side of the equation above comes
from computing the Euler-Lagrange equations of the NLO spin-orbit
potential \eqref{eq:25PNpot}, which was obtained in \cite{nrgrso}.
The result for this contribution can be conveniently arranged as
\begin{equation}
(\mathbf{a}_{1}^{i})^{V_{\textrm{SO}}^{\textrm{(2.5PN)}}}=\frac{1}{m_{1}}\sum_{n=0}^{3}\biggl\{(-1)^{n+1}\biggl(\frac{d}{dt}\biggr)^{n}\frac{\partial}{\partial\mathbf{x}_{1}^{i(n)}}V_{\text{SO}}^{\text{(2.5PN)}}\biggr\}=(\mathbf{A}_{1}^{i})_{S^{i0}}^{\textrm{cov}}+(\mathbf{A}_{1}^{i})_{S^{ij}},
\end{equation}
where
\begin{align}
(\mathbf{A}_{1}^{i})_{S^{i0}}^{\textrm{cov}} & \equiv\frac{G}{r^{3}}\biggl\{\frac{m_{2}}{m_{1}}S_{1}^{j0}\biggl[\delta^{ij}\biggl(\frac{Gm_{1}}{r}+2\frac{Gm_{2}}{r}+2\mathbf{v}\cdot\mathbf{v}_{2}+\frac{1}{2}\mathbf{a}_{2}\cdot\mathbf{r}+\frac{3}{2}(\mathbf{v}_{2}\cdot\mathbf{n})^{2}\biggr)+\mathbf{v}_{2}^{i}(3\mathbf{v}_{2}\cdot\mathbf{n}\mathbf{n}^{j}-\mathbf{v}^{j})+\frac{1}{2}\mathbf{a}_{2}^{i}\mathbf{r}^{j}\nonumber \\
 & \qquad+\mathbf{n}^{i}\biggl(-\frac{3}{2}r\mathbf{a}_{2}^{j}+3\mathbf{v}_{2}\cdot\mathbf{n}\mathbf{v}^{j}-\mathbf{n}^{j}\biggl(4\frac{Gm_{1}}{r}+8\frac{Gm_{2}}{r}+6\mathbf{v}\cdot\mathbf{v}_{2}+\frac{3}{2}\mathbf{a}_{2}\cdot\mathbf{r}+\frac{15}{2}(\mathbf{v}_{2}\cdot\mathbf{n})^{2}\biggr)\biggr)\biggr]\nonumber \\
 & \qquad+S_{2}^{j0}\biggl[\delta^{ij}\biggl(-2\frac{Gm_{1}}{r}-\frac{Gm_{2}}{r}+2\mathbf{v}\cdot\mathbf{v}_{1}+\frac{1}{2}\mathbf{a}_{1}\cdot\mathbf{r}-\frac{3}{2}(\mathbf{v}_{1}\cdot\mathbf{n})^{2}\biggr)-\mathbf{v}_{1}^{i}(3\mathbf{v}_{1}\cdot\mathbf{n}\mathbf{n}^{j}+\mathbf{v}^{j})+\frac{1}{2}\mathbf{a}_{1}^{i}\mathbf{r}^{j}\nonumber \\
 & \qquad+\mathbf{n}^{i}\biggl(-\frac{3}{2}r\mathbf{a}_{1}^{j}+3\mathbf{v}_{1}\cdot\mathbf{n}\mathbf{v}^{j}+\mathbf{n}^{j}\biggl(8\frac{Gm_{1}}{r}+4\frac{Gm_{2}}{r}-6\mathbf{v}\cdot\mathbf{v}_{1}-\frac{3}{2}\mathbf{a}_{1}\cdot\mathbf{r}+\frac{15}{2}(\mathbf{v}_{1}\cdot\mathbf{n})^{2}\biggr)\biggr)\biggr]\biggr\}\nonumber \\
 & -\frac{d}{dt}\biggl\{\frac{G}{r^{2}}\biggl[\frac{m_{2}}{m_{1}}S_{1}^{j0}(2\mathbf{v}_{2}^{i}\mathbf{n}^{j}-\delta^{ij}\mathbf{v}_{2}\cdot\mathbf{n})+S_{2}^{j0}[2\mathbf{n}^{j}(2\mathbf{v}_{1}^{i}-\mathbf{v}_{2}^{i})-\delta^{ij}\mathbf{v}_{1}\cdot\mathbf{n}-\mathbf{n}^{i}(\mathbf{v}^{j}+3\mathbf{v}_{1}\cdot\mathbf{n}\mathbf{n}^{j})]\biggr]\biggr\}\nonumber \\
 & +\frac{d^{2}}{dt^{2}}\biggl\{\frac{1}{2}\frac{G}{r}S_{2}^{j0}(3\delta^{ij}+\mathbf{n}^{i}\mathbf{n}^{j})\biggr\}
\end{align}
and
\begingroup
\allowdisplaybreaks
\begin{align}
(\mathbf{A}_{1}^{i})_{S^{ij}} & \equiv\frac{G}{r^{3}}\biggl\{\frac{m_{2}}{m_{1}}S_{1}^{ij}\biggl[-2\mathbf{v}_{2}\cdot\mathbf{r}\mathbf{a}_{2}^{j}-r^{2}\dot{\mathbf{a}}_{2}^{j}+\mathbf{v}_{1}^{j}\biggl(-\frac{Gm_{1}}{r}+\frac{1}{2}\frac{Gm_{2}}{r}+\frac{1}{2}\mathbf{a}_{2}\cdot\mathbf{r}+\frac{3}{2}(\mathbf{v}_{2}\cdot\mathbf{n})^{2}\biggl)\nonumber \\
 & \qquad+\mathbf{v}_{2}^{j}\biggl(-\frac{5}{2}\frac{Gm_{2}}{r}-2\mathbf{v}\cdot\mathbf{v}_{2}-\mathbf{a}_{2}\cdot\mathbf{r}-3(\mathbf{v}_{2}\cdot\mathbf{n})^{2}\biggr)\biggr]+S_{2}^{ij}\biggl[-2\mathbf{v}_{1}\cdot\mathbf{r}\mathbf{a}_{1}^{j}+r^{2}\dot{\mathbf{a}}_{1}^{j}\nonumber \\
 & \qquad+\mathbf{v}_{1}^{j}\biggl(\frac{5}{2}\frac{Gm_{1}}{r}-2\mathbf{v}\cdot\mathbf{v}_{1}-\mathbf{a}_{1}\cdot\mathbf{r}+3(\mathbf{v}_{1}\cdot\mathbf{n})^{2}\biggr)+\mathbf{v}_{2}^{j}\biggl(-\frac{1}{2}\frac{Gm_{1}}{r}+\frac{Gm_{2}}{r}+\frac{1}{2}\mathbf{a}_{1}\cdot\mathbf{r}-\frac{3}{2}(\mathbf{v}_{1}\cdot\mathbf{n})^{2}\biggr)\biggr]\nonumber \\
 & \qquad+\mathbf{n}^{i}\biggl[\frac{m_{2}}{m_{1}}S_{1}^{kj}\biggl(\biggl(-4\frac{Gm_{1}}{r}+2\frac{Gm_{2}}{r}+\frac{3}{2}\mathbf{a}_{2}\cdot\mathbf{r}+\frac{15}{2}(\mathbf{v}_{2}\cdot\mathbf{n})^{2}\biggr)\mathbf{v}_{1}^{k}\mathbf{n}^{j}-3\mathbf{v}_{2}\cdot\mathbf{n}(\mathbf{v}_{1}^{k}\mathbf{v}_{2}^{j}+2\mathbf{a}_{2}^{k}\mathbf{r}^{j})\nonumber \\
 & \qquad-\biggl(10\frac{Gm_{2}}{r}+6\mathbf{v}\cdot\mathbf{v}_{2}+3\mathbf{a}_{2}\cdot\mathbf{r}+15(\mathbf{v}_{2}\cdot\mathbf{n})^{2}\biggr)\mathbf{v}_{2}^{k}\mathbf{n}^{j}+\frac{1}{2}r\mathbf{a}_{2}^{k}\mathbf{v}_{1}^{j}+r\mathbf{a}_{2}^{k}\mathbf{v}_{2}^{j}+r\mathbf{r}^{k}\dot{\mathbf{a}}_{2}^{j}\biggr)\nonumber \\
 & \qquad+S_{2}^{kj}\biggl(\biggl(10\frac{Gm_{1}}{r}-6\mathbf{v}\cdot\mathbf{v}_{1}-3\mathbf{a}_{1}\cdot\mathbf{r}+15(\mathbf{v}_{1}\cdot\mathbf{n})^{2}\biggr)\mathbf{v}_{1}^{k}\mathbf{n}^{j}-3\mathbf{v}_{1}\cdot\mathbf{n}(\mathbf{v}_{1}^{k}\mathbf{v}_{2}^{j}+2\mathbf{a}_{1}^{k}\mathbf{r}^{j})\nonumber \\
 & \qquad+\biggl(-2\frac{Gm_{1}}{r}+4\frac{Gm_{2}}{r}+\frac{3}{2}\mathbf{a}_{1}\cdot\mathbf{r}-\frac{15}{2}(\mathbf{v}_{1}\cdot\mathbf{n})^{2}\biggr)\mathbf{v}_{2}^{k}\mathbf{n}^{j}+r\mathbf{a}_{1}^{k}\mathbf{v}_{1}^{j}+\frac{1}{2}r\mathbf{a}_{1}^{k}\mathbf{v}_{2}^{j}-r\mathbf{r}^{k}\dot{\mathbf{a}}_{1}^{j}\biggr)\biggr]\nonumber \\
 & \qquad+\mathbf{v}_{1}^{i}S_{2}^{kj}[\mathbf{v}_{1}^{k}\mathbf{v}_{2}^{j}+2\mathbf{a}_{1}^{k}\mathbf{r}^{j}+3\mathbf{v}_{1}\cdot\mathbf{n}(\mathbf{v}_{2}^{k}-2\mathbf{v}_{1}^{k})\mathbf{n}^{j}]+\mathbf{a}_{1}^{i}S_{2}^{kj}[\mathbf{v}_{1}^{k}\mathbf{r}^{j}-\frac{1}{2}\mathbf{v}_{2}^{k}\mathbf{r}^{j}]\nonumber \\
 & \qquad+\mathbf{v}_{2}^{i}\frac{m_{2}}{m_{1}}S_{1}^{kj}[\mathbf{v}_{1}^{k}\mathbf{v}_{2}^{j}+2\mathbf{a}_{2}^{k}\mathbf{r}^{j}+3\mathbf{v}_{2}\cdot\mathbf{n}(2\mathbf{v}_{2}^{k}-\mathbf{v}_{1}^{k})\mathbf{n}^{j}]+\mathbf{a}_{2}^{i}\frac{m_{2}}{m_{1}}S_{1}^{kj}\biggl[-\frac{1}{2}\mathbf{v}_{1}^{k}\mathbf{r}^{j}+\mathbf{v}_{2}^{k}\mathbf{r}^{j}\biggr]\biggr\}\nonumber \\
 & -\frac{d}{dt}\biggl\{\frac{G}{r^{2}}\biggl\{\frac{m_{2}}{m_{1}}S_{1}^{ij}\biggl[\frac{1}{2}r\mathbf{a}_{2}^{j}+\mathbf{v}_{2}\cdot\mathbf{n}\mathbf{v}_{2}^{j}+\mathbf{n}^{j}\biggl(\frac{Gm_{1}}{r}-\frac{1}{2}\frac{Gm_{2}}{r}-\frac{1}{2}\mathbf{a}_{2}\cdot\mathbf{r}-\frac{3}{2}(\mathbf{v}_{2}\cdot\mathbf{n})^{2}\biggr)\biggr]\nonumber \\
 & \qquad+S_{2}^{ij}\biggl[r\mathbf{a}_{1}^{j}+\mathbf{v}_{1}\cdot\mathbf{n}\mathbf{v}_{2}^{j}+\mathbf{n}^{j}\biggl(-\frac{5}{2}\frac{Gm_{1}}{r}+2\mathbf{v}\cdot\mathbf{v}_{1}+\mathbf{a}_{1}\cdot\mathbf{r}-3(\mathbf{v}_{1}\cdot\mathbf{n})^{2}\biggr)\biggr]+4\mathbf{v}_{1}^{i}S_{2}^{kj}\mathbf{v}_{1}^{k}\mathbf{n}^{j}\nonumber \\
 & \qquad+2\mathbf{v}_{2}^{i}\biggl[\frac{m_{2}}{m_{1}}S_{1}^{kj}\mathbf{v}_{2}^{k}\mathbf{n}^{j}-S_{2}^{kj}\mathbf{v}_{1}^{k}\mathbf{n}^{j}\biggr]+\mathbf{n}^{i}S_{2}^{kj}[\mathbf{v}_{1}^{k}\mathbf{v}_{2}^{j}+2\mathbf{a}_{1}^{k}\mathbf{r}^{j}+3\mathbf{v}_{1}\cdot\mathbf{n}\mathbf{n}^{j}(\mathbf{v}_{2}^{k}-2\mathbf{v}_{1}^{k})]\biggr\}\biggr\}\nonumber \\
 & +\frac{d^{2}}{dt^{2}}\biggl\{\frac{G}{r}[S_{2}^{ij}(2\mathbf{v}_{1}\cdot\mathbf{n}\mathbf{n}^{j}-\mathbf{v}_{1}^{j}-\frac{1}{2}\mathbf{v}_{2}^{j})+\mathbf{n}^{i}\mathbf{n}^{j}S_{2}^{kj}(\mathbf{v}_{1}^{k}-\frac{1}{2}\mathbf{v}_{2}^{k})]\biggr\}+\frac{d^{3}}{dt^{3}}\{GS_{2}^{ij}\mathbf{n}^{j}\}.
\end{align}
\endgroup
The second term in the right hand side of \eqref{eq:A1(2.5PN)gen}
accounts for 2.5PN order terms coming from order reduction of lower~PN order accelerations, which can be concisely presented as
\begin{align}
(\mathbf{a}_{1}^{i})^{\text{(Red.)}} & =\biggl[\frac{1}{2}\frac{Gm_{2}}{r}\mathbf{a}_{2}\cdot\mathbf{n}\mathbf{n}^{i}-\mathbf{a}_{1}\cdot\mathbf{v}_{1}\mathbf{v}_{1}^{i}-\mathbf{a}_{1}^{i}\biggl(3\frac{Gm_{2}}{r}+\frac{1}{2}\mathbf{v}_{1}^{2}\biggr)+\frac{7}{2}\frac{Gm_{2}}{r}\mathbf{a}_{2}^{i}\biggr]_{\mathbf{a}_{\text{SO}}^{\text{(1.5PN)}}}\nonumber \\
 & \quad+\left[\frac{G}{r^{3}}\biggl(-\frac{m_{2}}{m_{1}}S_{1}^{i0}+3\frac{m_{2}}{m_{1}}S_{1}^{j0}\mathbf{n}^{j}\mathbf{n}^{i}+S_{2}^{i0}-3S_{2}^{j0}\mathbf{n}^{j}\mathbf{n}^{i}\biggr)\right]_{\textrm{cov}\text{(1PN)}}\nonumber \\
 & \quad+\left[-\frac{G}{r^{3}}\biggl(\frac{m_{2}}{m_{1}}\dot{S}_{1}^{ij}\mathbf{r}^{j}+2\dot{S}_{2}^{ij}\mathbf{r}^{j}\biggr)\right]_{\dot{S}_{\text{LO}}}.
\end{align}
The expression above includes three contributions from lower-order
accelerations: reduced contributions from substituting the LO spin-orbit
acceleration \eqref{eq:a1(1.5PN)-1} in the acceleration terms present
in the 1PN correction to the equations of motion \eqref{eq:a1(1PN)};
frame corrections from imposing the covariant SSC \eqref{eq:CovSi0}
in \eqref{eq:a1(1.5PN)gen}, and also, in that same equation, terms
from reducing spin derivatives. At 2.5PN order, we only need the LO
spin derivative term given by \cite{nrgrs}
\begin{equation}
\frac{d\mathbf{S}_{1}}{dt}=\frac{Gm_{2}}{r^{3}}\bigl[2(\mathbf{r}\times\mathbf{v})\times\mathbf{S}_{1}+(\mathbf{S}_{1}\times\mathbf{r})\times\mathbf{v}_{1}\bigr].
\end{equation}

After imposing the covariant SSC and order reducing the accelerations
in order to obtain a fixed order result at 2.5PN, \eqref{eq:A1(2.5PN)gen}
becomes
\begingroup
\allowdisplaybreaks
\begin{align}
(\mathbf{a}_{1}^{i})_{\text{SO}}^{\text{(2.5PN)}} & =\frac{G}{r^{3}}\biggl\{-\frac{\mathbf{n}^{i}}{m\nu r}\biggl[\frac{m_{2}}{m_{1}}\mathbf{S}_{1}\cdot\mathbf{L}\biggl(\frac{G}{r}(26m_{1}+22m_{2})+12\mathbf{v}\cdot\mathbf{v}_{2}+3\mathbf{v}_{2}^{2}+3\mathbf{v}^{2}+15(\mathbf{v}_{2}\cdot\mathbf{n})^{2}\biggr)\nonumber \\
 & \qquad\qquad+\mathbf{S}_{2}\cdot\mathbf{L}\biggl(\frac{G}{r}\biggl(\frac{61}{2}m_{1}+20m_{2}\biggr)+6\mathbf{v}\cdot\mathbf{v}_{2}+3\mathbf{v}_{2}^{2}+15(\mathbf{v}_{2}\cdot\mathbf{n})^{2}\biggr)\biggr]\nonumber \\
 & \qquad\qquad+\mathbf{v}_{1}^{i}\biggl[-\frac{3m_{2}}{m_{1}}\biggl(\frac{1}{m\nu r}\mathbf{S}_{1}\cdot\mathbf{L}(2\mathbf{v}_{2}\cdot\mathbf{n}+\dot{r})+\dot{r}\mathbf{S}_{1}\cdot(\mathbf{v}_{2}\times\mathbf{n})+\mathbf{S}_{1}\cdot(\mathbf{v}\times\mathbf{v}_{2})\biggr)\nonumber \\
 & \qquad\qquad-2\biggl(\frac{3}{m\nu r}\mathbf{S}_{2}\cdot\mathbf{L}(\mathbf{v}_{2}\cdot\mathbf{n}+\dot{r})+3\dot{r}\mathbf{S}_{2}\cdot(\mathbf{v}_{2}\times\mathbf{n})+2\mathbf{S}_{2}\cdot(\mathbf{v}\times\mathbf{v}_{2})\biggr)\biggr]\nonumber \\
 & \qquad\qquad+\mathbf{v}_{2}^{i}\biggl[\frac{6}{m\nu r}\biggl(\frac{m_{2}}{m_{1}}\mathbf{S}_{1}+\mathbf{S}_{2}\biggr)\cdot\mathbf{L}(\mathbf{v}_{2}\cdot\mathbf{n}+\dot{r})\biggr]-\frac{2}{m\nu r}\mathbf{L}^{i}\biggl[\frac{G}{r}\biggl(\frac{m_{2}^{2}}{m_{1}}\mathbf{S}_{1}\cdot\mathbf{n}+2m_{1}\mathbf{S}_{2}\cdot\mathbf{n}\biggr)\biggr]\nonumber \\
 & \qquad\qquad+\frac{m_{2}}{m_{1}}(\mathbf{S}_{1}\times\mathbf{n})^{i}\biggl[\dot{r}\frac{G}{r}(14m_{1}+10m_{2})+\frac{3}{2}\dot{r}\bigl(\mathbf{v}_{1}^{2}+5(\mathbf{v}_{2}\cdot\mathbf{n})^{2}\bigr)-3\mathbf{v}\cdot\mathbf{v}_{2}\mathbf{v}_{2}\cdot\mathbf{n}\biggr]\nonumber \\
 & \qquad\qquad+(\mathbf{S}_{2}\times\mathbf{n})^{i}\biggl[\dot{r}\frac{G}{r}\biggl(\frac{47}{2}m_{1}+16m_{2}\biggr)-2\mathbf{v}_{2}\cdot\mathbf{n}\biggl(\frac{Gm_{1}}{r}+3\mathbf{v}\cdot\mathbf{v}_{2}\biggr)+3\dot{r}\bigl(2\mathbf{v}\cdot\mathbf{v}_{2}+\mathbf{v}_{2}^{2}+5(\mathbf{v}_{2}\cdot\mathbf{n})^{2}\bigr)\biggr]\nonumber \\
 & \qquad\qquad-\frac{m_{2}}{m_{1}}(\mathbf{S}_{1}\times\mathbf{v})^{i}\biggl[\frac{G}{r}(14m_{1}+10m_{2})+6\mathbf{v}\cdot\mathbf{v}_{2}+\frac{3}{2}\mathbf{v}_{2}^{2}+\frac{3}{2}\mathbf{v}^{2}+\frac{9}{2}(\mathbf{v}_{2}\cdot\mathbf{n})^{2}-3\dot{r}\mathbf{v}_{2}\cdot\mathbf{n}\biggr]\nonumber \\
 & \qquad\qquad-(\mathbf{S}_{2}\times\mathbf{v})^{i}\biggl[\frac{G}{r}\biggl(\frac{31}{2}m_{1}+12m_{2}\biggr)+4\mathbf{v}\cdot\mathbf{v}_{2}+2\mathbf{v}_{2}^{2}+6(\mathbf{v}_{2}\cdot\mathbf{n})^{2}\biggr]\biggr\}\label{eq:Am1m2}.
\end{align}
\endgroup
Note that the spin vector used in this expression is defined in the
locally flat frame; see section \ref{sec:Comparison} for a discussion
of alternative spin definitions.

We also present the NLO spin-orbit acceleration in the center-of-mass
frame. In the latter, the expressions for $\mathbf{x}_{1}$ and $\mathbf{x}_{2}$
in terms of the relative coordinate\textbf{ $\mathbf{r}$} are given
by
\begin{align}
\mathbf{x}_{1} & =\frac{m_{2}}{m}\mathbf{r}+\delta\mathbf{r},\label{eq:cm115}\\
\mathbf{x}_{2} & =-\frac{m_{1}}{m}\mathbf{r}+\delta\mathbf{r},\label{eq:x2cm}
\end{align}
where, considering only corrections up to 1.5PN order,
\begin{equation}
\delta\mathbf{r}_{\text{}}=\nu\frac{\delta m}{2m}\biggl(\mathbf{v}^{2}-\frac{Gm}{r}\biggr)\mathbf{r}+\frac{\nu}{m}\mathbf{v}\times\mathbf{\Sigma}.\label{eq:delta_rcm}
\end{equation}
When these PN corrections to the center-of-mass frame are considered
in the 1PN acceleration \eqref{eq:a1(1PN)}, they yield contributions
to the equations of motion at the 2.5PN order. In principle, one must
consider 1PN center-of-mass corrections in the LO spin-orbit acceleration
\eqref{eq:a1(1.5PN)-1} as well, but these vanish because this acceleration
only depends on relative coordinates and velocities; this is also
the reason why we do not need to consider 2.5PN spin-orbit correction
to the center-of-mass in the Newtonian acceleration \eqref{eq:a1(0PN)}.
Therefore, the final expression for the NLO spin-orbit acceleration
in the center-of-mass frame comes solely from considering \eqref{eq:cm115}
and \eqref{eq:x2cm} in \eqref{eq:Am1m2} and \eqref{eq:a1(1PN)};
the result is
\begingroup
\allowdisplaybreaks
\begin{align}
(\mathbf{a}^{i})_{\text{SO}}^{\text{(2.5PN)}}= & \frac{G}{m\nu r^{4}}\biggl\{\mathbf{n}^{i}\biggl[\mathbf{S}\cdot\mathbf{L}\biggl(-\frac{Gm}{r}(42+29\nu)+3(-1+10\nu)\mathbf{v}^{2}-30\nu\dot{r}^{2}\biggr)\nonumber \\
 & \qquad\quad-\frac{\delta m}{m}\boldsymbol{\Sigma}\cdot\mathbf{L}\biggl(\frac{Gm}{r}\biggl(22+\frac{33}{2}\nu\biggr)+3(1-5\nu)\mathbf{v}^{2}+15\nu\dot{r}^{2}\biggr)\biggr]\nonumber \\
 & \qquad\quad+3\dot{r}\mathbf{v}^{i}\biggl[3\mathbf{S}\cdot\mathbf{L}(-1+\nu)+\frac{\delta m}{m}\boldsymbol{\Sigma}\cdot\mathbf{L}(-1+2\nu)\biggr]\nonumber \\
 & \qquad\quad-2\frac{Gm}{r}\mathbf{L}^{i}\biggl[\mathbf{S}\cdot\mathbf{n}(1+2\nu)+\frac{\delta m}{m}\boldsymbol{\Sigma}\cdot\mathbf{n}(1+\nu)\biggr]\biggr\}\nonumber \\
 & +\frac{G}{r^{3}}\biggl\{(\mathbf{S}\times\mathbf{n})^{i}\dot{r}\biggl[\frac{Gm}{r}(26+25\nu)+\frac{3}{2}(1-15\nu)\mathbf{v}^{2}+\frac{45}{2}\nu\dot{r}^{2}\biggr]\nonumber \\
 & \qquad\quad+\frac{\delta m}{m}(\boldsymbol{\Sigma}\times\mathbf{n})^{i}\dot{r}\biggl[\frac{Gm}{r}\biggl(10+\frac{27}{2}\nu\biggr)+\biggl(\frac{3}{2}-12\nu\biggr)\mathbf{v}^{2}+15\nu\dot{r}^{2}\biggr]\nonumber \\
 & \qquad\quad+(\mathbf{S}\times\mathbf{v})^{i}\biggl[-\frac{Gm}{r}(22+15\nu)+\frac{3}{2}(-1+11\nu)\mathbf{v}^{2}-\frac{33}{2}\nu\dot{r}^{2}\biggr]\nonumber \\
 & \qquad\quad-\frac{\delta m}{m}(\boldsymbol{\Sigma}\times\mathbf{v})^{i}\biggl[\frac{Gm}{r}\biggl(10+\frac{15}{2}\nu\biggr)+\biggl(\frac{3}{2}-8\nu\biggr)\mathbf{v}^{2}+9\nu\dot{r}^{2}\biggr]\biggr\}.\label{eq:Acm}
\end{align}
\endgroup
This expression is valid for general orbits and for arbitrary spin
orientations within the region of validity of the NRGR formalism.
In the next section, we compute the binding energy and the energy
loss. For the latter, we need the result \eqref{eq:Acm} as well as
\eqref{eq:a1(0PN)}, \eqref{eq:a1(1PN)}, and \eqref{eq:a1(1.5PN)-1}
to order reduce the time derivatives of the multipole moments.

\section{Binding energy and energy loss\label{sec:Energy}}

\subsection{Binding Energy}

The LO spin-orbit energy--a 1.5PN correction to the Newtonian binding
energy--can be obtained from the potential \eqref{eq:15PNpot}; it
is given by
\begin{equation}
E_{\text{SO}}^{\text{(1.5PN)}}=\frac{G}{r^{3}}\mathbf{r}^{i}(m_{2}S_{1}^{i0}-m_{1}S_{2}^{i0})_{\text{cov}}=\frac{Gm_{2}}{r^{2}}\mathbf{S}_{1}\cdot(\mathbf{n}\times\mathbf{v}_{1})+1\leftrightarrow2,\label{eq:ELOSO}
\end{equation}
where we have imposed the covariant SSC in the second expression.
In this section, we obtain the 1PN correction to the LO spin-orbit
binding energy:
\begin{align}
E_{\text{SO}}^{\text{(2.5PN)}} & =\sum_{A=1}^{2}\sum_{n=0}^{2}\mathbf{p}_{\mathbf{x}_{A}^{(n)}}\cdot\mathbf{x}_{A}^{(n+1)}+V_{\text{SO}}^{\text{(2.5PN)}}+E^{\text{(Red.)}},\label{eq:E_hamil}\\
\mathbf{p}_{q^{(n)}} & =-\sum_{A=1}^{2}\sum_{k=n+1}^{3}\biggl(-\frac{d}{dt}\biggr)^{k-n-1}\frac{\partial V_{\text{SO}}^{\text{(2.5PN)}}}{\partial\mathbf{x}_{A}^{(k)}},
\end{align}
where the notation $\mathbf{x}_{A}^{(k)}$ is a compact way to express
$\frac{d^{k}\mathbf{x}_{A}}{dt^{k}}$. For the spin-orbit energy at
the 2.5PN order, we have two contributions: one from the NLO spin-orbit
potential \eqref{eq:25PNpot}, and another from frame corrections
when applying the covariant SSC to the LO spin-orbit energy \eqref{eq:ELOSO},
which we represent by $E^{\text{(Red.)}}$ in \eqref{eq:E_hamil}.
The sum of the two contributions gives
\begin{align}
E_{\text{SO}}^{\text{(2.5PN)}} & =\frac{Gm_{2}}{r^{2}}\biggl\{(\mathbf{v}_{1}\cdot\mathbf{n}+2\mathbf{v}_{2}\cdot\mathbf{n})\mathbf{S}_{1}\cdot(\mathbf{v}_{1}\times\mathbf{v}_{2})\nonumber \\
 & \qquad+\biggl(\mathbf{v}_{2}^{2}-\mathbf{v}_{1}\cdot\mathbf{v}_{2}-\frac{3}{2}(\mathbf{v}_{2}\cdot\mathbf{n})^{2}-3\mathbf{v}_{1}\cdot\mathbf{n}\mathbf{v}_{2}\cdot\mathbf{n}+2\frac{Gm_{1}}{r}\biggr)\mathbf{S}_{1}\cdot(\mathbf{v}_{1}\times\mathbf{n})\nonumber \\
 & \qquad+\biggl(-2\mathbf{v}_{2}^{2}+3\mathbf{v}_{1}\cdot\mathbf{v}_{2}-\mathbf{v}_{1}^{2}+3(\mathbf{v}_{1}\cdot\mathbf{n})^{2}+3\mathbf{v}_{1}\cdot\mathbf{n}\mathbf{v}_{2}\cdot\mathbf{n}+3\frac{Gm_{2}}{r}\biggr)\mathbf{S}_{1}\cdot(\mathbf{v}_{2}\times\mathbf{n})\biggr\}+1\leftrightarrow2.\label{eq:Em1m2}
\end{align}
 Transforming to the center-of-mass frame as in section \ref{sec:EOM},
we have
\begin{equation}
E_{\text{SO}}^{\text{(2.5PN)}}=\frac{Gm}{r}\biggl\{\biggl(2\nu\frac{Gm}{r}-2\mathbf{v}^{2}-\frac{3}{2}\nu\dot{r}^{2}\biggr)\frac{\mathbf{L}\cdot\mathbf{S}}{mr^{2}}+\frac{\delta m}{m}\biggl(\frac{3}{2}\nu\frac{Gm}{r}-\frac{3}{2}\nu\mathbf{v}^{2}\biggr)\frac{\mathbf{L}\cdot\mathbf{\Sigma}}{mr^{2}}\biggr\}.\label{eq:Ecm}
\end{equation}
Next we calculate the time-averaged energy loss, which completes the
pieces necessary to compute the orbital phase evolution in section
\ref{sec:Phase}.

\subsection{Energy Loss\label{subsec:Energy-Loss}}

The binary system's energy loss due to the emission of graviational
waves can be computed directly from the one-graviton emission amplitude
in the effective theory (for a detailed discussion, see \cite{andirad,andirad2}),
and its general form is given at \eqref{eq:dEdt}. All the necessary
multipole moments to compute the NLO spin-orbit effects in the energy
loss are presented in (\ref{eq:mass_quad01}--\ref{eq:curr_octo_spin}).
Using the equations of motion \eqref{eq:A_PNexpanded} to order reduce
the acceleration terms generated by the time derivatives applied to
the multipole moments, we obtain a final expression for the NLO spin-orbit
energy loss:
\begin{align}
\frac{dE}{dt}\bigg|_{\text{SO}}^{\text{(2.5PN)}}= & -\frac{2G^{3}m^{3}\nu}{105r^{4}}\biggl\{\frac{\mathbf{L}\cdot\mathbf{S}}{mr^{2}}\biggl[(3776+1560\nu)\frac{G^{2}m^{2}}{r^{2}}+(-12892+2024\nu)\frac{Gm}{r}\dot{r}^{2}+(15164-560\nu)\frac{Gm}{r}\mathbf{v}^{2}\nonumber \\
 & \qquad\qquad\quad+(-8976+12576\nu)\dot{r}^{4}+(13362-18252\nu)\dot{r}^{2}\mathbf{v}^{2}+(-4226+5952\nu)\mathbf{v}^{4}\biggr]\nonumber \\
 & \qquad\quad+\frac{\delta m}{m}\frac{\mathbf{L}\cdot\mathbf{\Sigma}}{mr^{2}}\biggl[(-548+952\nu)\frac{G^{2}m^{2}}{r^{2}}+(-14654+4796\nu)\frac{Gm}{r}\dot{r}^{2}+(10718-1708\nu)\frac{Gm}{r}\mathbf{v}^{2}\nonumber \\
 & \qquad\qquad\quad+(-7941+10704\nu)\dot{r}^{4}+(8742-13434\nu)\dot{r}^{2}\mathbf{v}^{2}+(-2001+3474\nu)\mathbf{v}^{4}\biggr]\biggr\}.\label{eq:Pcm}
\end{align}
Along with the expressions for the acceleration \eqref{eq:Acm} and
conserved energy \eqref{eq:Ecm}, the above result is the final piece
needed to compute the orbital phase evolution of the binary system,
in the quasi-circular orbit approximation, accounting for NLO spin-orbit
effects in the NRGR framework.

\section{Phase evolution\label{sec:Phase}}

Until this point, our results are valid for general orbits and arbitrary
spin configurations. However, as is well known, the emission of gravitational
waves tends to efficiently circularize orbits well before entering
the observable frequency band of gravitational wave detectors \cite{LincolnCircOrbs}.
Although alternative methods such as the dynamical renormalization
group approach \cite{chadDRG,zixinDRG} may be used for more general
systems, we will restrict our analysis to circular orbits here. We
can then apply an adiabatic approximation in which orbits are approximately
circular on a orbital time scale and orbit decay occurs on a radiation-reaction
time scale. In this approximation, the expressions above can be expressed
as coordinate-independent quantities as functions of a single orbital
frequency $\omega$, the orbital angular momentum $L$, and the spin
vectors, and are gauge invariant under coordinate transformations.
In our subsequent analysis, we neglect spin-spin \cite{nrgrs}, tail
\cite{andirad,rec1_Galley_2016} and radiation-reaction \cite{chadbr1,chadbr2}
terms in the orbital frequency, since in this paper we are only investigating
spin-orbit effects; those other effects do not mix with our results
and thus can be included independently later on.

For non-spinning objects, the procedure of computing the phase evolution
of the binary system is unambiguous because the orientation of the
orbital plane is constant in time; for spinning objects, the choice
of spin vector is crucial because the spin vector may evolve by radiation
reaction for an inappropriate choice. For spinning systems, we choose
a spin vector with conserved norm; this allows us to work with orbit
averaged spin vectors and to use energy balance arguments to compute
the orbital phase \cite{Will1,luc13}. We transform from the locally
flat spin vectors to conserved norm spin vectors using the relation
\begin{equation}
\mathbf{S}_{A}\rightarrow\biggl(1+\frac{1}{2}\mathbf{v}_{A}^{2}\biggr)\mathbf{S}_{A}^{c}-\frac{1}{2}\mathbf{v}_{A}(\mathbf{S}_{A}^{c}\cdot\mathbf{v}_{A})+\cdots.\label{eq:Sc-1-1}
\end{equation}
See \cite{nrgrso} and section \ref{sec:Comparison} for details regarding
the significance of this redefinition.

For (quasi-)circular orbits, we use the relations
\begin{align}
r\omega^{2} & =-\langle\mathbf{n}\cdot\mathbf{a}\rangle,\label{eq:omega2}\\
|\mathbf{v}| & =r\omega,\\
\dot{r} & =0,
\end{align}
and perform the spin transformation to conserved norm spin vectors,
which gives us, for instance,
\begin{equation}
E_{\text{SO}}^{c} =\frac{G}{r^{3}}\biggl\{\biggl[1+2\nu\frac{Gm}{r}-\frac{3}{2}(1+\nu)\mathbf{v}^{2}\biggr]\mathbf{L}\cdot\mathbf{S}^{c}+\frac{\delta m}{m}\biggl[1+\frac{3}{2}\nu\frac{Gm}{r}+\frac{1}{2}(1-5\nu)\mathbf{v}^{2}\biggr]\mathbf{L}\cdot\mathbf{\Sigma}^{c}\biggr\}\label{eq:Ec}
\end{equation}
and
\begin{align}
\frac{dE^{c}}{dt}\bigg|_{\text{SO}} & =\nu\frac{G^{3}m^{2}}{105r^{6}}\biggl\{\mathbf{L}\cdot\mathbf{S}^{c}\biggl[448\frac{Gm}{r}+4480\mathbf{v}^{2}\nonumber \\
 & \qquad\qquad\qquad-(7552+3120\nu)\frac{G^{2}m^{2}}{r^{2}}-(30440+1792\nu)\frac{Gm}{r}\mathbf{v}^{2}+(9656-14480\nu)\mathbf{v}^{4}\biggr]\nonumber \\
 & \qquad\qquad+\frac{\delta m}{m}\mathbf{L}\cdot\mathbf{\Sigma}^{c}\biggl[-224\frac{Gm}{r}+2408\mathbf{v}^{2}\nonumber \\
 & \qquad\qquad\qquad+(1906-1904\nu)\frac{G^{2}m^{2}}{r^{2}}-(21548-3976\nu)\frac{Gm}{r}\mathbf{v}^{2}+(5206-8320\nu)\mathbf{v}^{4}\biggr]\biggr\}.\label{eq:Pc}
\end{align}
To write these in terms of the orbital frequency $\omega$, we use
equation \eqref{eq:omega2} and solve order by order in the PN expansion
for $\omega$; note that the expression for the acceleration \eqref{eq:Acm}
must also be rewritten with the conserved norm spin vectors. Then,
we find that the orbital frequency is given by
\begin{align}
\omega^{2} & =\frac{Gm}{r^{3}}\biggl\{1+\frac{Gm}{r}(-3+\nu)-\biggl(\frac{Gm}{r}\biggr)^{9/2}\biggl[5\frac{S_{\ell}^{c}}{Gm^{2}}+3\frac{\delta m}{m}\frac{\Sigma_{\ell}^{c}}{Gm^{2}}\biggr]\nonumber \\
 & \qquad\qquad\qquad+\biggl(\frac{Gm}{r}\biggr)^{2}\biggl[\frac{41}{4}\nu+\nu^{2}\Bigr]+\biggl(\frac{Gm}{r}\biggr)^{11/2}\biggl[\biggl(\frac{27}{2}-\frac{13}{2}\nu\biggr)\frac{\delta m}{m}\frac{\Sigma_{\ell}^{c}}{Gm^{2}}+\biggl(\frac{45}{2}-\frac{27}{2}\nu\biggr)\frac{S_{\ell}^{c}}{Gm^{2}}\biggr]\biggr\}+\cdots,
\end{align}
where $S_{\ell}^{c}\equiv\mathbf{\hat{\ell}}\cdot\mathbf{S}^{c}$,
$\Sigma_{\ell}^{c}\equiv\mathbf{\hat{\ell}}\cdot\mathbf{\Sigma}^{c}$,
and $\hat{\ell}=\mathbf{L}/|\mathbf{L}|$ . We can write equations
\eqref{eq:Ec}, \eqref{eq:Pc} in terms of the orbital separation~$r$ to give
\begin{align}
E^{c}(r) & =-\frac{1}{2}\frac{Gm^{2}\nu}{r}\biggl\{1+\frac{Gm}{r}\biggl[-\frac{7}{4}+\frac{1}{4}\nu\biggr]+\biggl(\frac{Gm}{r}\biggr)^{3/2}\biggl[\frac{\delta m}{m}\frac{\Sigma_{\ell}^{c}}{Gm^{2}}+3\frac{S_{\ell}^{c}}{Gm^{2}}\biggr]\nonumber \\
 & \qquad\qquad\qquad+\biggl(\frac{Gm}{r}\biggr)^{2}\biggl[-\frac{23}{8}+\frac{49}{8}\nu+\frac{1}{8}\nu^{2}\biggr]+\biggl(\frac{Gm}{r}\biggr)^{5/2}\biggl[(2-3\nu)\frac{\delta m}{m}\frac{\Sigma_{\ell}^{c}}{Gm^{2}}+(6-6\nu)\frac{S_{\ell}^{c}}{Gm^{2}}\biggr]\biggr\}
 \end{align}
 and
 \begin{align}
\frac{dE^{c}(r)}{dt} & =-\frac{32G^{4}m^{5}\nu^{2}}{5r^{5}}\biggl\{1+\frac{Gm}{r}\biggl[-\frac{2927}{336}-\frac{5}{4}\nu\biggr]+\biggl(\frac{Gm}{r}\biggr)^{3/2}\biggl[-\frac{25}{4}\frac{\delta m}{m}\frac{\Sigma_{\ell}^{c}}{Gm^{2}}-\frac{37}{3}\frac{S_{\ell}^{c}}{Gm^{2}}\biggr]\nonumber \\
 & \qquad+\biggl(\frac{Gm}{r}\biggr)^{2}\biggl[\frac{202663}{9072}+\frac{380}{9}\nu\biggr]+\biggl(\frac{Gm}{r}\biggr)^{5/2}\biggl[\biggl(\frac{6953}{112}+\frac{91}{8}\nu\biggr)\frac{\delta m}{m}\frac{\Sigma_{\ell}^{c}}{Gm^{2}}+\biggl(\frac{18947}{168}+\frac{68}{3}\nu\biggr)\frac{S_{\ell}^{c}}{Gm^{2}}\biggr]\biggr\}.
\end{align}

These two expressions depend on the coordinate separation $r$, and
are therefore gauge dependent. Inverting our expression for $\omega^{2}$,
we find
\begin{align}
\frac{Gm}{r} & =x+x^{2}\biggl[1-\frac{1}{3}\nu\biggr]+\frac{x^{5/2}}{Gm^{2}}\biggl[\frac{\delta m}{m}\Sigma_{\ell}^{c}+\frac{5}{3}S_{\ell}^{c}\biggr]+x^{3}\biggl[3-\frac{65}{12}\nu\biggr]+\frac{x^{7/2}}{Gm^{2}}\biggl[2\frac{\delta m}{m}\Sigma_{\ell}^{c}+\biggl(\frac{10}{3}+\frac{8}{9}\nu\biggr)S_{\ell}^{c}\biggr],
\end{align}
where the PN parameter $x\equiv(Gm\omega)^{2/3}$ is formally of order
$v^{2}$. We can now write the energy and energy loss as gauge independent
expressions. They are
\begin{align}
E^{c}(x) & =-\frac{1}{2}m\nu x\biggl\{1+x\biggl[-\frac{3}{4}-\frac{1}{12}\nu\biggr]+\frac{x^{3/2}}{Gm^{2}}\biggl[2\frac{\delta m}{m}\Sigma_{\ell}^{c}+\frac{14}{3}S_{\ell}^{c}\biggr]\nonumber \\
 & \qquad\qquad\qquad\qquad\qquad+x^{2}\biggl[-\frac{27}{8}+\frac{19}{8}\nu-\frac{1}{24}\nu^{2}\biggr]+\frac{x^{5/2}}{Gm^{2}}\biggl[\biggl(3-\frac{10}{3}\nu\biggl)\frac{\delta m}{m}\Sigma_{\ell}^{c}+\biggl(11-\frac{61}{9}\nu\biggr)S_{\ell}^{c}\biggr]\biggl\}\label{eq:Ecx}
 \end{align}
 and
 \begin{align}
\frac{dE^{c}(x)}{dt} & =-\frac{32x^{5}\nu^{2}}{5G}\biggl\{1+x\biggl[-\frac{1247}{336}-\frac{35}{12}\nu\biggr]+\frac{x^{3/2}}{Gm^{2}}\biggl[-\frac{5}{4}\frac{\delta m}{m}\Sigma_{\ell}^{c}-4S_{\ell}^{c}\biggr]\nonumber \\
 & \qquad\qquad\qquad+x^{2}\biggl[-\frac{44711}{9072}+\frac{9271}{504}\nu+\frac{65}{18}\nu^{2}\biggr]+\frac{x^{5/2}}{Gm^{2}}\biggl[\biggl(-\frac{13}{16}+\frac{43}{4}\nu\biggr)\frac{\delta m}{m}\Sigma_{\ell}^{c}+\biggl(-\frac{9}{2}+\frac{272}{9}\nu\biggr)S_{\ell}^{c}\biggr]\biggr\}.\label{eq:Pcx}
\end{align}
The coefficients in these expressions are still dependent on the particular
definition of the spins; with our choice of conserved norm spin vectors,
equations \eqref{eq:Ecx} and \eqref{eq:Pcx} yield perfect agreement
with the corresponding expressions in \cite{buo2}. We now proceed
to find an expression for the phase evolution of the binary system
using energy balance arguments. We first obtain a dimensionless adiabatic
parameter (also called the orbital frequency evolution \cite{kidder})
representing the orbital decay, given by
\begin{align}
\frac{\dot{\omega}}{\omega^{2}}=\frac{96}{5}\nu x^{5/2} & \biggl\{1+x\biggl[-\frac{743}{336}-\frac{11}{4}\nu\biggr]+\frac{x^{3/2}}{Gm^{2}}\biggl[-\frac{25}{4}\frac{\delta m}{m}\Sigma_{\ell}^{c}-\frac{47}{3}S_{\ell}^{c}\biggr]\nonumber \\
 & +x^{2}\biggl[\frac{34103}{18144}+\frac{13661}{2016}\nu+\frac{59}{18}\nu^{2}\biggr]+\frac{x^{5/2}}{Gm^{2}}\biggl[\biggl(-\frac{809}{84}+\frac{281}{8}\nu\biggr)\frac{\delta m}{m}\Sigma_{\ell}^{c}+\biggl(-\frac{5861}{144}+\frac{1001}{12}\nu\biggr)S_{\ell}^{c}\biggr]\biggr\}.\label{eq:adiabatic}
\end{align}
The orbital phase can then be computed in this adiabatic approximation,
where the gravitational wave phase contains two contributions. The
first comes from the evolution of the carrier phase, while the second
arises due to the precession of the orbital plane due to spin effects.
This can schematically be written $\Phi_{\text{GW}}=\phi_{\text{GW}}+\delta\phi$
using the notation of \cite{buo2}. The carrier phase given by $\phi_{\text{GW}}=2\phi$
can be computed using
\begin{align}
\phi & =\int dt\:\omega=\int d\omega\,\frac{\omega}{\dot{\omega}}.
\end{align}
In general, the carrier phase may be computed numerically for arbitrary
spin alignments. However, for spins aligned or anti-aligned with the
binary orbital angular momentum, this can be computed analytically
using equation \eqref{eq:adiabatic} to yield
\begin{align}
\phi & =\phi_{0}-\frac{32}{\nu}\biggl\{ x^{-5/2}+x^{-3/2}\biggl[\frac{3715}{1008}+\frac{55}{12}\nu\biggr]+\frac{x^{-1}}{Gm^{2}}\biggl[\frac{125}{8}\frac{\delta m}{m}\Sigma_{\ell}^{c}+\frac{235}{6}S_{\ell}^{c}\biggr]\nonumber \\
 & \qquad+x^{-1/2}\biggl[\frac{15293365}{1016064}+\frac{27145}{1008}\nu+\frac{3085}{144}\nu^{2}\biggr]-\frac{\text{log\,}x}{Gm^{2}}\biggl[\biggl(\frac{41745}{448}-\frac{15}{8}\nu\biggr)\frac{\delta m}{m}\Sigma_{\ell}^{c}+\biggl(\frac{554345}{2016}+\frac{55}{8}\nu\biggr)S_{\ell}^{c}\biggr]\biggr\},\label{eq:orbital_phase}
\end{align}
for which we find perfect agreement with \cite{buo2}.

\section{Center-of-mass correction\label{sec:CM}}

We now proceed to compute the center-of-mass correction at 2.5PN order
due to NLO spin-orbit effects. But before proceeding to the details
of its computation, notice that this correction should have, in principle,
entered in the calculation of the quantities derived in the previous
sections, namely the NLO spin-orbit acceleration \eqref{eq:Acm},
binding energy \eqref{eq:Ecm} and the energy loss \eqref{eq:Pcm}.
The reason why this correction does not affect the result for the
NLO spin-orbit acceleration, as previously explained in section \ref{sec:EOM},
is that the Newtonian acceleration \eqref{eq:a1(0PN)} is naturally
given in terms of relative coordinates. This argument does not hold
for the Newtonian energy, but it turns out that the 2.5PN contribution
that would arise from it cancels out due to its symmetry:
\begin{align}
E^{\text{(0PN)}} & =\frac{m_{1}\mathbf{v}_{1}^{2}}{2}+\frac{m_{2}\mathbf{v}_{2}^{2}}{2}-\frac{Gm_{1}m_{2}}{r}\xrightarrow{2.5PN}\frac{m_{1}}{2}\frac{m_{2}}{m}\mathbf{v}\cdot\delta\dot{\mathbf{r}}_{\text{SO}}^{\text{(2.5PN)}}-\frac{m_{2}}{2}\frac{m_{1}}{m}\mathbf{v}\cdot\delta\dot{\mathbf{r}}_{\text{SO}}^{\text{(2.5PN)}}=0.
\end{align}

The same happens to the LO mass quadrupole moment $I_{0PN}^{ij}=\sum_{a}m_{a}\left[\mathbf{x}_{a}^{i}\mathbf{x}_{a}^{j}\right]_{TF}$
when we try to extract its 2.5PN contribution going to the center-of-mass
frame, and consequently the energy loss due to NLO spin-orbit effects
is not affected by the correction to the center-of-mass at this order.
Despite of these facts, the NLO spin-orbit correction to the center-of-mass,
which is an effect that enters at 2.5PN order, itself is a non-zero
quantity and must be obtained, since it will lead to non-zero contributions
in future computations at N$^{2}$LO order. Below, we present how
we proceed to obtain this quantity via the NRGR framework.

The center-of-mass position is defined as
\begin{equation}
\mathbf{r}_{\mathrm{cm}}^{i}=\frac{1}{m}\int d^{3}x\,\mathbf{x}^{i}\,T^{00}(\mathbf{x},t).\label{eq:xcmT}
\end{equation}
As previously mentioned in section \ref{subsec:Radiative-sector},
we can extract the stress-energy pseudotensor $T^{\mu\nu}(\mathbf{x},t)$
from matching onto the effective action \eqref{eq:Sonegrav-1-1} by
integrating out potential modes from the full theory action in equation
\eqref{eq:full}. Introducing the partial Fourier transform of the
stress-energy pseudotensor and taking the long-wavelength limit, we
find
\begin{align}
T^{\mu\nu}(\mathbf{q},t) & =\int d^{3}x\,T^{\mu\nu}(\mathbf{x},t)e^{-i\mathbf{q}\cdot\mathbf{x}}\\
 & =\ensuremath{\sum_{n=0}^{\infty}\frac{(-i)^{n}}{n!}\biggl(\int d^{3}xT^{\mu\nu}(\mathbf{x},t)\mathbf{x}^{i_{1}}\ldots\mathbf{x}^{i_{n}}\biggr)\mathbf{q}_{i_{1}}\ldots\mathbf{q}_{i_{n}}}.\label{eq:Tmunuq}
\end{align}
Comparing equations \eqref{eq:xcmT} and \eqref{eq:Tmunuq}, we can
read off the center-of-mass correction from the $\mathcal{O}(\mathbf{q})$
term in $T^{\mu\nu}(\mathbf{q},t)$ in the effective theory.

\begin{figure}
\begin{centering}
\subfloat[]{\includegraphics[scale=0.25]{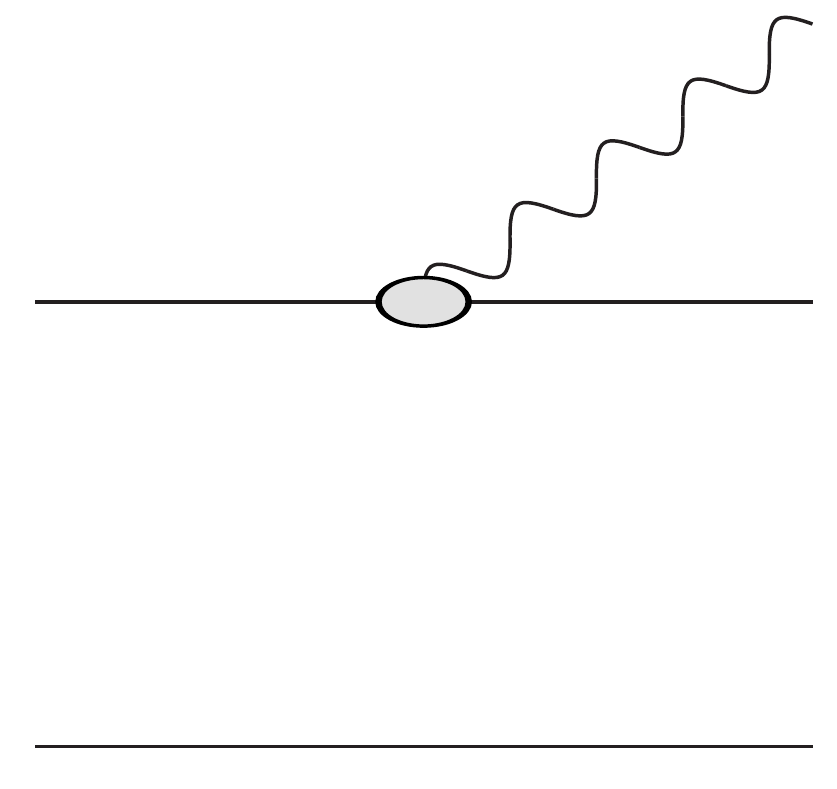}

}\hspace{4em}\subfloat[]{\includegraphics[scale=0.25]{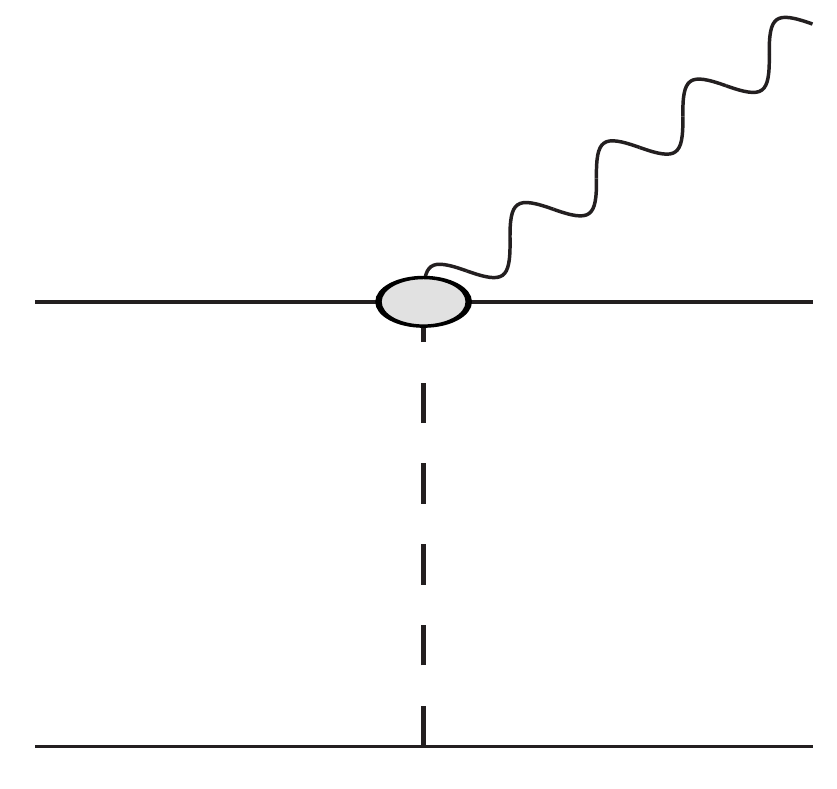}

}\hspace{4em}\subfloat[]{\includegraphics[scale=0.25]{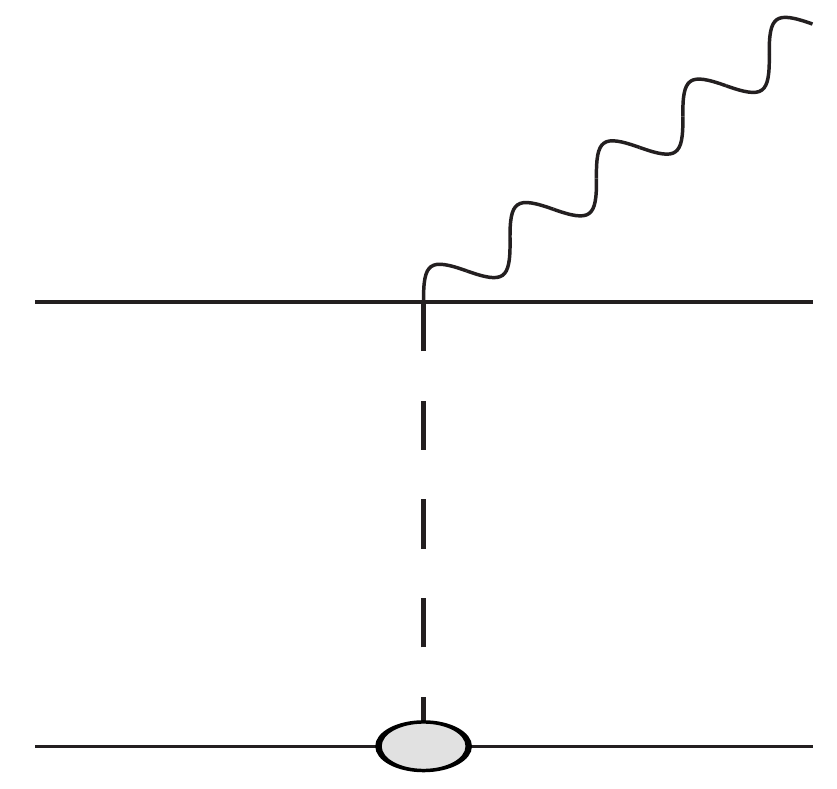}

}\hspace{4em}\subfloat[]{\includegraphics[scale=0.25]{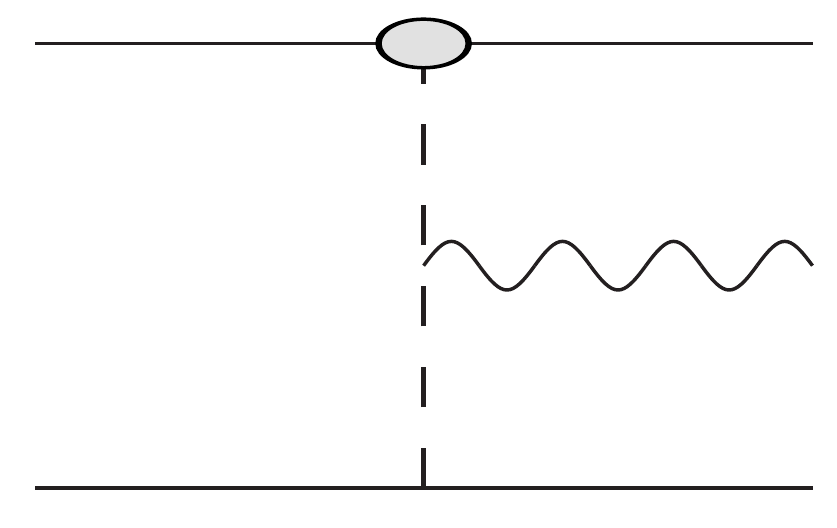}

}
\par\end{centering}
\caption{Diagrams contributing to the 2.5PN spin-orbit center-of-mass correction.}
\label{fig_cm}
\end{figure}
The diagrams that contribute to the NLO spin-orbit center-of-mass
correction are given in figure \ref{fig_cm}. Diagram \ref{fig_cm}a
comes from a single insertion of the vertex \eqref{eq:hSi0v1}. Imposing
the covariant SSC gives a LO spin-orbit term and a 1PN correction
given by
\begin{equation}
T_{\mathrm{\text{\ref{fig_cm}a}}}^{00}(t,\mathbf{q})=\sum_{A\neq B}S_{A}^{0i}(i\mathbf{q}^{i})e^{-i\mathbf{q}\cdot\mathbf{x}_{A}}\xrightarrow{(\textrm{cov})}\sum_{A\neq B}S_{A}^{ij}(i\mathbf{q}^{j})\biggl(\mathbf{v}_{A}^{i}+\frac{2Gm_{B}}{r}\mathbf{v}^{i}\biggr)e^{-i\mathbf{q}\cdot\mathbf{x}_{A}}.
\end{equation}

At the order we are working, figure \ref{fig_cm}b is composed of
two different contributions, as we show next. Contracting \eqref{eq:HhSijv0}
with \eqref{eq:Hv1}, we find
\begin{equation}
T_{\mathrm{\text{\ref{fig_cm}b,1}}}^{00}(t,\mathbf{q})=\sum_{A\neq B}\biggl[-\frac{2G_{N}m_{B}}{r}S_{A}^{ij}\mathbf{v}_{B}^{i}(i\mathbf{q}^{j})\biggr]e^{-i\mathbf{q}\cdot\mathbf{x}_{A}},
\end{equation}
and contracting \eqref{eq:HhSi0v1} with \eqref{eq:Hv0} gives
\begin{equation}
T_{\mathrm{\text{\ref{fig_cm}b,2}}}^{00}(t,\mathbf{q})=\sum_{A\neq B}\frac{Gm_{B}S_{A}^{0j}\mathbf{r}^{j}}{r^{3}}e^{-i\mathbf{q}\cdot\mathbf{x}_{A}}.
\end{equation}

Figure \ref{fig_cm}c also accounts for two distinct contributions.
Contracting \eqref{eq:HSv0} with \eqref{eq:Hhv1}, we have
\begin{equation}
T_{\mathrm{\text{\ref{fig_cm}c,1}}}^{00}(t,\mathbf{q})=\sum_{A\neq B}\frac{2Gm_{A}S_{B}^{ij}\mathbf{v}_{A}^{i}\mathbf{r}^{j}}{r^{3}}e^{-i\mathbf{q}\cdot\mathbf{x}_{A}},
\end{equation}
and contracting \eqref{eq:HSv1} with \eqref{eq:Hhv0} gives
\begin{equation}
T_{\mathrm{\text{\ref{fig_cm}c,2}}}^{00}(t,\mathbf{q})=\sum_{A\neq B}\biggl[\frac{Gm_{A}}{r^{3}}(-S_{B}^{0i}\mathbf{r}^{i}-S_{B}^{ij}\mathbf{v}_{B}^{i}\mathbf{r}^{j})\biggr]e^{i\mathbf{q}\cdot\mathbf{x}_{A}}.
\end{equation}

Finally, figure \ref{fig_cm}d comes from three different contractions.
The first contribution, constructed from \eqref{eq:Hv0} and \eqref{eq:HSv1}
together with the LO 3-point vertex gives
\begin{align}
T_{\mathrm{\text{\ref{fig_cm}d,1}}}^{00}(t,\mathbf{q}) & =\sum_{A\neq B}\biggl[\frac{3}{2}\frac{Gm_{A}}{r^{3}}S_{B}^{0j}\mathbf{r}^{j}-\frac{3}{2}\frac{Gm_{B}}{r^{3}}S_{A}^{0j}\mathbf{r}^{j}-\frac{3}{2}\frac{Gm_{B}}{r}S_{A}^{0j}(i\mathbf{q}^{j})\nonumber \\
 & \quad\qquad\hfill+\frac{3}{2}\frac{Gm_{A}}{r^{3}}S_{B}^{ij}\mathbf{v}_{B}^{i}\mathbf{r}^{j}+\frac{1}{2}\frac{Gm_{B}}{r^{3}}S_{A}^{ij}\mathbf{v}_{A}^{i}\mathbf{r}^{j}+\frac{1}{2}\frac{Gm_{B}}{r}S_{A}^{ij}\mathbf{v}_{A}^{i}(i\mathbf{q}^{j})\biggr]e^{-i\mathbf{q}\cdot\mathbf{x}_{A}}.
\end{align}
The second, constructed from \eqref{eq:Hv1} and \eqref{eq:HSv0}
with the LO 3-point vertex, is
\begin{equation}
T_{\mathrm{\text{\ref{fig_cm}d,2}}}^{00}(t,\mathbf{q})=\sum_{A\neq B}\biggl[-\frac{Gm_{A}}{r^{3}}S_{B}^{ij}\mathbf{v}_{A}^{i}\mathbf{r}^{j}+\frac{Gm_{B}}{r^{3}}S_{A}^{ij}\mathbf{v}_{B}^{i}\mathbf{r}^{j}+\frac{Gm_{B}}{r}S_{A}^{ij}\mathbf{v}_{B}^{i}(i\mathbf{q}^{j})\biggr]e^{-i\mathbf{q}\cdot\mathbf{x}_{A}}.
\end{equation}
The third, constructed from \eqref{eq:Hv0} and \eqref{eq:HSv0} with
the 3-point vertex at $\mathcal{O}(v^{1})$ reads
\begin{align}
T_{\mathrm{\text{\ref{fig_cm}d,3}}}^{00}(t,\mathbf{q}) & =\sum_{A\neq B}\biggl[\frac{Gm_{A}}{r}S_{B}^{ij}(\mathbf{v}_{A}^{i}+\mathbf{v}_{B}^{i})(i\mathbf{q}^{j})-\frac{Gm_{A}}{r^{3}}S_{B}^{ij}\mathbf{r}^{i}(i\mathbf{q}^{j})\mathbf{r}\cdot(\mathbf{v}_{A}+\mathbf{v}_{B})\biggr]e^{-i\mathbf{q}\cdot\mathbf{x}_{A}}.
\end{align}

Now, putting all the contributions above together, we write the final
expression for the $00$-component of the stress-pseudo tensor accounting
for NLO spin-orbit terms:
\begin{align}
T_{\text{SO}}^{00}(t,\mathbf{q}) & =\sum_{A\neq B}\biggl\{ S_{A}^{0i}(i\mathbf{q}^{i})+\frac{G}{r^{3}}\biggl[\tfrac{1}{2}m_{A}S_{B}^{0j}\mathbf{r}^{j}-\tfrac{1}{2}m_{B}S_{A}^{0j}\mathbf{r}^{j}+m_{B}(\mathbf{v}_{B}^{i}+\tfrac{1}{2}\mathbf{v}_{A}^{i})\mathbf{r}^{j}S_{A}^{ij}+m_{A}(\tfrac{1}{2}\mathbf{v}_{B}^{i}+\mathbf{v}_{A}^{i})\mathbf{r}^{j}S_{B}^{ij}\nonumber \\
 & \qquad\qquad+\bigl(-\tfrac{3}{2}m_{B}S_{A}^{0j}+m_{B}(\tfrac{1}{2}\mathbf{v}_{A}^{i}-\mathbf{v}_{B}^{i})S_{A}^{ij}+m_{A}(\mathbf{v}_{A}^{k}+\mathbf{v}_{B}^{k})(\delta^{ik}-\mathbf{n}^{i}\mathbf{n}^{k})S_{B}^{ij}\bigr)r^{2}(i\mathbf{q}^{j})\biggr]\biggr\} e^{-i\mathbf{q}\cdot\mathbf{x}_{A}}.
\end{align}

We can extract some information regarding the binary system from the
expression above when we take the long-wavelength limit by Taylor
expanding it around $\mathbf{q}=0$. For instance, the zeroth order
terms in the Taylor expansion give us the LO spin-orbit energy
\begin{equation}
E_{\text{SO}}^{\text{(1.5PN)}}=\int d^{3}x\,T^{00}(\mathbf{x},t)=-\sum_{A\neq B}\frac{Gm_{B}}{r^{3}}S_{A}^{0j}\mathbf{r}^{j},\label{eq:eso}
\end{equation}
and this serves as a self-consistency check, since \eqref{eq:eso}
agrees with equation \eqref{eq:ELOSO}, which we calculated from the
LO spin-orbit potential. Next, the terms linear in $\mathbf{q}$ yield
the center-of-mass position \eqref{eq:xcmT}, which is also conveniently
expressed through\footnote{The expression for $\mathbf{G}$ can be expanded order by order as
$\mathbf{G}=\mathbf{G}^{\text{(0PN)}}+\mathbf{G}^{\text{(1PN)}}+\mathbf{G}_{\text{SO}}^{\text{(1.5PN)}}+\mathbf{G}^{(\text{2PN)}}+\mathbf{G}^{(\text{2.5PN})}+\cdots$;
the LO and 1PN corrections can be found in \cite{andirad}, while
the 2PN correction was computed in \cite{2PN_paper}.} $\mathbf{G}\equiv m\mathbf{r}_{\text{cm}}$:
\begin{align}
\mathbf{G}_{\text{(1.5PN)}}^{k} & =-\sum_{A=1}^{2}S_{A}^{0k}=-\sum_{A=1}^{2}S_{A}^{ik}\mathbf{v}_{A}^{i},\\
\mathbf{G}_{\text{(2.5PN)}}^{k} & =\sum_{A\neq B}\frac{Gm_{B}}{r^{3}}\Bigl[S_{A}^{ij}\mathbf{r}^{j}(\mathbf{v}_{B}^{i}\mathbf{r}^{k}-\mathbf{v}_{A}^{i}\mathbf{x}_{B}^{k})-S_{A}^{ik}\bigl(2r^{2}\mathbf{v}^{i}-\mathbf{r}^{i}\mathbf{r}\cdot(\mathbf{v}_{A}+\mathbf{v}_{B})\bigr)\Bigr].
\end{align}

Now, in order to extract its corrections, we put the center-of-mass
at the origin, meaning $\mathbf{G}=0$, and iteratively solve for
$\mathbf{x}_{1}$, $\mathbf{x}_{2}$. Writing
\begin{align}
\mathbf{x}_{1}= & \frac{m_{2}}{m}\mathbf{r}+\delta\mathbf{r}^{\text{(1PN)}}+\delta\mathbf{r}_{\text{SO}}^{\text{(1.5PN)}}+\delta\mathbf{r}^{(\text{2PN})}+\delta\mathbf{r}_{\text{SO}}^{(\text{2.5PN})}+\cdots,\\
\mathbf{x}_{2}= & -\frac{m_{1}}{m}\mathbf{r}+\delta\mathbf{r}^{\text{(1PN)}}+\delta\mathbf{r}_{\text{SO}}^{\text{(1.5PN)}}+\delta\mathbf{r}^{(\text{2PN})}+\delta\mathbf{r}_{\text{SO}}^{(\text{2.5PN})}+\cdots,
\end{align}
we can determine PN corrections to the center-of-mass order by order.
The corrections $\delta\mathbf{r}^{\text{(1PN)}}$ and $\delta\mathbf{r}_{\text{SO}}^{\text{(1.5PN)}}$
can be found in \cite{andirad,kidder} and are presented in equation
\eqref{eq:cm115}, while the non-spin\footnote{There is no spin correction to the center-of-mass position at 2PN
order.} $\delta\mathbf{r}^{(\text{2PN})}$ can be found in \cite{2PN_paper}.
The NLO correction, with covariant SSC enforced, is
\begin{align}
\delta\mathbf{r}_{\text{SO}}^{\text{(2.5PN)}} & =\frac{\nu}{2m}\biggl\{\biggl[\nu\mathbf{v}^{2}-\frac{Gm}{r}(4+2\nu)\biggr]\mathbf{\Sigma}\times\mathbf{v}+\frac{\delta m}{m}\biggl[\mathbf{v}^{2}-\frac{Gm}{r}\biggr]\mathbf{S}\times\mathbf{v}\nonumber \\
 & \qquad\qquad+\frac{2Gm}{r}\biggl[\frac{\delta m}{m}\mathbf{S}\cdot(\mathbf{v}\times\mathbf{n})\mathbf{n}+\frac{3}{2}\frac{\delta m}{m}\dot{r}(\mathbf{S}\times\mathbf{n})+(1-4\nu)\dot{r}(\mathbf{\Sigma}\times\mathbf{n})\biggr]\biggr\}.\label{eq:COMNLO}
\end{align}

\section{Correspondence with other formalisms\label{sec:Comparison}}

At this point, we note that the expressions for the acceleration \eqref{eq:Acm},
the binding energy \eqref{eq:Ecm}, the NLO spin-orbit multipole moments
(\ref{eq:Iij_NLOSO}, \ref{eq:Jij_NLOSO}), and the center-of-mass
correction \eqref{eq:COMNLO} take a different form than the corresponding
results given in the literature \cite{buo1,buo2,bohennloso,luc13}.
As emphasized throughout this paper, we work with spins defined in
the locally flat frame. We would expect, then, that an appropriate
spin transformation coupled with a coordinate transformation should
give agreement with existing results; the difficulty reduces to finding
the appropriate set of transformations. As was discussed in \cite{nrgrso},
it is possible to construct an equivalent Hamiltonian to those in
\cite{buo1,damournloso} and thus the equations of motion were expected
to agree. In particular, there are two sets of results we would like
show agreement with: those for spin written in the PN frame as in
\cite{buo1,buo2}, and those with spins of constant magnitude as in
\cite{bohennloso,luc13}.

The relationship between the locally flat spin vectors and the PN
spin vectors was shown in \cite{nrgrss}. In the locally flat frame,
we chose the relation between the spin tensor and spin vector in \eqref{eq:CovSij}.
A natural definition of the spin tensor in terms of the spin vector
in the PN frame is
\begin{equation}
S^{\mu\nu}=-\frac{1}{m\sqrt{-g}}\epsilon^{\mu\nu\rho\sigma}p_{\rho}S_{\sigma},\label{eq:tensortovec-1}
\end{equation}
which clearly preserves the covariant SSC, and which in the locally
flat frame reduces to \eqref{eq:CovSij}. We fix the spin vector by
imposing the additional condition used in \cite{owen} given by
\begin{equation}
S^{\mu}p_{\mu}=0.
\end{equation}
From these definitions, it was shown in \cite{nrgrss} that the transformation
from the locally flat spin vectors to the PN spin vectors $\mathbf{\bar{S}}_{A}$
to 1PN order is given by
\begin{align}
\mathbf{S}_{A} & \rightarrow\biggl(1+\frac{\mathbf{v}_{A}^{2}}{2}+\frac{Gm_{B}}{r}\biggr)\bar{\mathbf{S}}_{A}-\mathbf{v}_{A}(\mathbf{\bar{S}}_{A}\cdot\mathbf{v}_{A}).
\end{align}
This transformation induces a 1PN correction to the spins, and was
used in that paper to show equivalence between the spin evolution
equations in \cite{nrgrss} and \cite{buo1,owen}. Note that to leading
order in the spins, the locally flat and PN frames are equivalent;
corrections only enter at 1PN order. For NLO spin-orbit effects, there
is a contribution that leads to different expressions for the accelerations,
energy, mass quadrupole, current quadrupole, and energy loss and center-of-mass
correction. With this spin transformation, the acceleration \eqref{eq:Acm},
the binding energy \eqref{eq:Ecm}, the multipole moments (\ref{eq:Iij_NLOSO},
\ref{eq:Jij_NLOSO}), the energy loss \eqref{eq:Pcm}, and the center-of-mass
correction \eqref{eq:COMNLO} agree completely with the corresponding
results in \cite{buo1,buo2}. Importantly, the general expressions
for these quantities agree exactly even \emph{before} writing gauge
invariant quantities. Of particular interest, the multipole moments
agree completely with those in \cite{buo2}, showing that the EFT
formalism used in this paper agrees with the literature, when spin-orbit
effects are considered beyond the dominant order, not only in the
conservative but also in the dissipative sector.

We also present the transformation to constant magnitude spin vectors
as used in computing the orbital phase \eqref{eq:orbital_phase}.
This spin choice was used in \cite{buo2,bohennloso,luc13}, and as
discussed in section \ref{sec:Phase} is the proper choice when computing
quantities in the adiabatic approximation. As shown in \cite{nrgrso},
the transformation to 1PN order is given by \eqref{eq:Sc-1-1}. This
puts the spin evolution equations into a spin precession form \cite{nrgrso,buo1},
i.e.,
\begin{equation}
\frac{d\mathbf{S}_{A}^{c}}{dt}=\mathbf{\Omega}_{A}\times\mathbf{S}_{A}^{c},
\end{equation}
where $\mathbf{\Omega}_{A}$ is the precessional frequency. This spin
transformation takes us from the covariant SSC to the Newton--Wigner
SSC, with one important caveat. Completing the transformation to the
Newton--Wigner SSC requires a change of coordinates that accounts
for the shift in the center-of-mass of each binary consituent (see
\cite{Barker1974,kidder,nrgrso} for a detailed discussion). In fact,
this spin redefinition coupled with the coordinate transformation
to the Newton--Wigner SSC is the only possible choice if one wants
to work with canonical variables \cite{SteinhoffCanSpin}. However,
to show the equivalence between our results and those in the literature,
we forego the coordinate transformation and find that our results
for the acceleration \eqref{eq:Acm}, the binding energy \eqref{eq:Ecm},
the multipole moments (\ref{eq:Iij_NLOSO}, \ref{eq:Jij_NLOSO}),
and the center-of-mass correction \eqref{eq:COMNLO} agree completely
with the corresponding results in \cite{buo2,bohennloso,luc13} with
conserved norm spins.

\section{Final remarks\label{sec:Final-remarks}}

We used the potential obtained in \cite{nrgrso} via the NRGR formalism
\cite{nrgr,nrgrs} to compute the NLO spin-orbit correction to the
equations of motion and to the binding energy of a binary system of
compact bodies in its inspiral stage. This correction to the equation
of motion, which is a 2.5PN acceleration, was used together with the
multipole moments computed in \cite{Porto:2010zg} to calculate the
NLO spin-orbit terms in the energy lost by the system due to the emission
of gravitational waves. Then, we utilized these results to compute
the evolution of the orbital frequency and, consequently, of the orbital
phase of the binary system accounting for spin-orbit effects beyond
the dominant order, considering quasi-circular orbits within the adiabatic
approximation. In performing these computations, we have made extensive
use of the \emph{Mathematica} package xAct \cite{xact}. In addition,
we calculated the 2.5PN spin-orbit terms of the $00$-component of
the pseudotensor of the system in order to extract the correction
to the center-of-mass associated to NLO spin-orbit effects.

Although the results of this paper--the NLO spin-orbit effects in
the equations of motion, center-of-mass frame, binding energy, energy
loss, orbital evolution and phase evolution--only now were obtained
in the NRGR framework, they had been previously computed through other
formalisms that follow more conventional approaches to general relativity.
Therefore, we provided a discussion in which we explained that our
EFT results and those found in the literature \cite{buo1,buo2,bohennloso,luc13}
are in perfect agreement once appropriate spin transformations are
considered. While the equivalence between the EFT formalism and other
methods was demonstrated in \cite{nrgrso} in the conservative sector
regarding NLO spin-orbit effects, we have shown now full agreement
also in the radiation sector.

Moreover, while inviting for the completion of higher order spin computations,
the results obtained in this paper provide the final missing pieces
needed to compute waveforms that include subleading spin-orbit effects
entirely within the NRGR formalism, which will be presented in a future
publication.

\section{Acknowledgements}

We thank Rafael Porto for the useful discussions on the subjects presented
in this paper. We also thank Adam Leibovich for the valuable suggestions
in the preparation of this manuscript. B.P. and N.T.M. are supported
in part by the National Science Foundation under Grant No. PHY-1820760.

\appendix

\section{Toolkit\label{sec:App-A}}

\subsection*{Non-spin accelerations}

The PN corrections to the Newtonian acceleration of one of the bodies--let
us choose body 1--in the binary system are given below. In the EFT
formalism, the 1PN correction to the LO gravitational acceleration
\begin{equation}
(\mathbf{a}_{1}^{i})_{\text{}}^{\text{(0PN)}}=-\frac{Gm_{2}}{r^{2}}\mathbf{n}^{i},\label{eq:a1(0PN)}
\end{equation}
can be derived from the Lagrangian obtained in \cite{nrgr}, and it
reads as
\begin{align}
(\mathbf{a}_{1}^{i})^{\text{(1PN)}}= & \frac{Gm_{2}}{2r^{2}}\biggl\{\mathbf{n}^{i}\biggl[\frac{2Gm}{r}-3(\mathbf{v}_{1}^{2}+\mathbf{v}_{2}^{2})+7\mathbf{v}_{1}\cdot\mathbf{v}_{2}+3\mathbf{v}_{1}\cdot\mathbf{n}\mathbf{v}_{2}\cdot\mathbf{n}\biggr]-\mathbf{v}_{2}\cdot\mathbf{n}\mathbf{v}_{1}^{i}-\mathbf{v}_{1}\cdot\mathbf{n}\mathbf{v}_{2}^{i}+\dot{r}(6\mathbf{v}_{1}^{i}-7\mathbf{v}_{2}^{i}-\mathbf{n}^{i}\mathbf{v}_{2}\cdot\mathbf{n})\nonumber \\
 & \qquad\qquad-6r\mathbf{a}_{1}^{i}+7r\mathbf{a}_{2}^{i}+(\mathbf{v}^{i}-\mathbf{n}^{i}\dot{r})\mathbf{v}_{2}\cdot\mathbf{n}+r\mathbf{a}_{2}\cdot\mathbf{n}\mathbf{n}^{i}+\mathbf{n}^{i}\bigl(\mathbf{v}_{2}\cdot(\mathbf{v}-\mathbf{n}\dot{r})\bigr)\biggr\}-\frac{1}{2}\mathbf{a}_{1}^{i}\mathbf{v}_{1}^{2}-\mathbf{v}_{1}^{i}\mathbf{v}_{1}\cdot\mathbf{a}_{1}.\label{eq:a1(1PN)}
\end{align}

The second PN correction to the gravitational acceleration was derived
in \cite{2PN_paper} considering the EFT theory in the linearized
harmonic gauge, and it is given as follows:
\begin{align}
(\mathbf{a}_{1}^{i})_{}^{\text{(2PN)}} & =\frac{1}{8}\frac{Gm_{2}}{r^{3}}\mathbf{r}^{i}\biggl\{\frac{G^{2}}{r^{2}}(-2m_{1}^{2}-20m_{1}m_{2}+16m_{2}^{2})+\frac{G}{r}\biggl[(18m_{1}+56m_{2})\mathbf{v}_{1}^{2}\nonumber \\
 & \qquad-(84m_{1}+128m_{2})\mathbf{v}_{1}\cdot\mathbf{v}_{2}+(58m_{1}+64m_{2})\mathbf{v}_{2}^{2}+30m_{1}\mathbf{a}_{1}\cdot\mathbf{r}-12m\mathbf{a}_{2}\cdot\mathbf{r}\nonumber \\
 & \qquad+\frac{28}{r^{2}}(m_{1}-4m_{2})\mathbf{v}_{1}\cdot\mathbf{r}(\mathbf{v}_{1}\cdot\mathbf{r}-2\mathbf{v}_{2}\cdot\mathbf{r})-\frac{1}{r^{2}}(56m_{1}+176m_{2})(\mathbf{v}_{2}\cdot\mathbf{r})^{2}\biggr]\nonumber \\
 & \qquad+2\mathbf{v}_{1}^{4}-16(\mathbf{v}_{1}\cdot\mathbf{v}_{2})^{2}-16\mathbf{v}_{2}^{4}+32\mathbf{v}_{1}\cdot\mathbf{v}_{2}\mathbf{v}_{2}^{2}-2\mathbf{v}_{1}^{2}\mathbf{a}_{2}\cdot\mathbf{r}-2\mathbf{v}_{2}^{2}\mathbf{a}_{2}\cdot\mathbf{r}\nonumber \\
 & \qquad-4\mathbf{a}_{2}\cdot\mathbf{v}_{2}\mathbf{v}_{2}\cdot\mathbf{r}+\frac{(\mathbf{v}_{2}\cdot\mathbf{r})^{2}}{r^{2}}(12\mathbf{v}_{1}^{2}-48\mathbf{v}_{1}\cdot\mathbf{v}_{2}+36\mathbf{v}_{2}^{2})-15\frac{(\mathbf{v}_{2}\cdot\mathbf{r})^{4}}{r^{4}}\biggr\}\nonumber \\
 & +\frac{1}{4}\frac{Gm_{2}}{r^{3}}\mathbf{v}_{1}^{i}\biggl\{\frac{G}{r}\biggl[(48m_{2}-15m_{1})\mathbf{v}_{1}\cdot\mathbf{r}+(23m_{1}-40m_{2})\mathbf{v}_{2}\cdot\mathbf{r}\biggr]\nonumber \\
 & \qquad+\mathbf{v}_{2}\cdot\mathbf{r}(4\mathbf{v}_{1}^{2}+16\mathbf{v}_{1}\cdot\mathbf{v}_{2}-20\mathbf{v}_{2}^{2})-24\frac{\mathbf{v}_{1}\cdot\mathbf{r}(\mathbf{v}_{2}\cdot\mathbf{r})^{2}}{r^{2}}+18\frac{(\mathbf{v}_{2}\cdot\mathbf{r})^{3}}{r^{2}}\nonumber \\
 & \qquad+\mathbf{v}_{1}\cdot\mathbf{r}(8\mathbf{v}_{1}^{2}-16\mathbf{v}_{1}\cdot\mathbf{v}_{2}+16\mathbf{v}_{2}^{2}-2\mathbf{a}_{2}\cdot\mathbf{r})+2r^{2}(12\mathbf{a}_{1}-7\mathbf{a}_{2})\cdot\mathbf{v}_{1}\biggr\}\nonumber \\
 & +2\mathbf{a}_{1}\cdot\mathbf{v}_{1}\mathbf{v}_{1}^{2}\mathbf{v}_{1}^{i}+\frac{1}{4}\mathbf{a}_{1}^{i}\biggl(49\frac{G^{2}m_{1}m_{2}}{r^{2}}+36\frac{G^{2}m_{2}^{2}}{r^{2}}+12\frac{Gm_{2}}{r}\mathbf{v}_{1}^{2}+\mathbf{v}_{1}^{4}\biggr)\nonumber \\
 & +\frac{1}{4}\frac{Gm_{2}}{r^{3}}\mathbf{v}_{2}^{i}\biggl\{\frac{G}{r}\biggl[(31m_{1}-24m_{2})\mathbf{v}_{1}\cdot\mathbf{r}+(40m_{2}-9m_{1})\mathbf{v}_{2}\cdot\mathbf{r}\biggr]\nonumber \\
 & \qquad+\mathbf{v}_{2}\cdot\mathbf{r}(-4\mathbf{v}_{1}^{2}-16\mathbf{v}_{1}\cdot\mathbf{v}_{2}+20\mathbf{v}_{2}^{2})+24\frac{\mathbf{v}_{1}\cdot\mathbf{r}(\mathbf{v}_{2}\cdot\mathbf{r})^{2}}{r^{2}}-18\frac{(\mathbf{v}_{2}\cdot\mathbf{r})^{3}}{r^{2}}\nonumber \\
 & \qquad+\mathbf{v}_{1}\cdot\mathbf{r}(16\mathbf{v}_{1}\cdot\mathbf{v}_{2}-16\mathbf{v}_{2}^{2})-14r^{2}\mathbf{a}_{2}\cdot\mathbf{v}_{2}\biggr\}-\frac{7}{4}\frac{Gm_{2}}{r}\mathbf{a}_{2}^{i}\biggl(6\frac{Gm}{r}+\mathbf{v}_{1}^{2}+\mathbf{v}_{2}^{2}\biggr).\label{eq:a1(2PN)}
\end{align}

\subsection*{Spin-orbit potentials}

The LO and NLO spin-orbit potentials \cite{nrgrs,nrgrso}--from which
the LO and NLO spin-orbit accelerations and binding energies are computed--read,
respectively, as
\begin{align}
V_{\mathrm{SO}}^{\mathrm{(1.5PN)}}= & \frac{G\mathbf{r}^{j}}{r^{3}}\biggl\{ m_{2}(S_{1}^{j0}+S_{1}^{jk}\mathbf{v}_{1}^{k}-2S_{1}^{jk}\mathbf{v}_{2}^{k})-m_{1}(S_{2}^{j0}+S_{2}^{jk}\mathbf{v}_{2}^{k}-2S_{2}^{jk}\mathbf{v}_{1}^{k})\biggr\},\label{eq:15PNpot}\\
V_{\mathrm{SO}}^{\mathrm{2.5PN}}= & \frac{Gm_{2}}{r^{3}}\biggl\{\biggl[S_{1}^{i0}\biggl(2\mathbf{v}_{2}^{2}-2\mathbf{v}_{1}\cdot\mathbf{v}_{2}-\frac{3}{2r^{2}}(\mathbf{v}_{2}\cdot\mathbf{r})^{2}-\frac{1}{2}\mathbf{a}_{2}\cdot\mathbf{r}\biggr)\nonumber \\
 & \qquad\quad+\biggl(2\mathbf{v}_{1}\cdot\mathbf{v}_{2}+\frac{3(\mathbf{v}_{2}\cdot\mathbf{r})^{2}}{r^{2}}-2\mathbf{v}_{2}^{2}+\mathbf{a}_{2}\cdot\mathbf{r}\biggr)S_{1}^{ij}\mathbf{v}_{2}^{j}\nonumber \\
 & \qquad\quad-\biggl(\frac{3}{2r^{2}}(\mathbf{v}_{2}\cdot\mathbf{r})^{2}+\frac{1}{2}\mathbf{a}_{2}\cdot\mathbf{r}\biggr)S_{1}^{ij}\mathbf{v}_{1}^{j}+2S_{1}^{ij}\mathbf{a}_{2}^{j}\mathbf{v}_{2}\cdot\mathbf{r}+r^{2}S_{1}^{ij}\dot{\mathbf{a}}_{2}^{j}\biggr]\mathbf{r}^{i}\nonumber \\
 & \qquad\quad+S_{1}^{i0}\biggl((\mathbf{v}_{1}-\mathbf{v}_{2})^{i}\mathbf{v}_{2}\cdot\mathbf{r}-\frac{3}{2}\mathbf{a}_{2}^{i}r^{2}\biggr)+S_{1}^{ij}\biggl(\mathbf{v}_{2}^{i}\mathbf{v}_{1}^{j}\mathbf{v}_{2}\cdot\mathbf{r}-r^{2}\mathbf{a}_{2}^{j}\mathbf{v}_{2}^{i}-\frac{1}{2}r^{2}\mathbf{a}_{2}^{j}\mathbf{v}_{1}^{i}\biggr)\biggr\}\nonumber \\
 & +\frac{G^{2}m_{2}}{r^{4}}\mathbf{r}^{i}\biggl[-(m_{1}+2m_{2})S_{1}^{i0}+\biggl(m_{1}-\frac{m_{2}}{2}\biggr)S_{1}^{ij}\mathbf{v}_{1}^{j}+\frac{5m_{2}}{2}S_{1}^{ij}\mathbf{v}_{2}^{j}\biggr]+1\leftrightarrow2.\label{eq:25PNpot}
\end{align}

\subsection*{Multipole moments}

The multipole moments needed to compute the energy loss at 2.5PN were
obtained in \cite{andirad,Porto:2010zg}. We present them here, written
in the center-of-mass frame and with the covariant SSC imposed. The
spin vector is defined in the locally flat frame. The mass quadrupole
moments are
\begin{align}
I_{\text{(0PN)}}^{ij} & =m\nu\{\mathbf{r}^{i}\mathbf{r}^{j}\}_{\text{TF}},\label{eq:mass_quad01}\\
I_{\text{(1PN)}}^{ij} & =m\nu\biggl\{\biggl[\biggl(-\frac{5}{7}+\frac{8}{7}\nu\biggr)\frac{Gm}{r}+\biggl(\frac{29}{42}-\frac{29}{14}\nu\biggr)\mathbf{v}^{2}\biggr]\mathbf{r}^{i}\mathbf{r}^{j}+\biggl(\frac{11}{21}-\frac{11}{7}\nu\biggr)r^{2}\mathbf{v}^{i}\mathbf{v}^{j}+\biggl(-\frac{4}{7}+\frac{12}{7}\nu\biggr)r\dot{r}\mathbf{v}^{j}\mathbf{r}^{i}\biggr\}_{\text{STF}},\\
I_{\text{(1.5PN)}}^{ij} & =\nu\biggl\{\frac{8}{3}(\mathbf{v}\times\mathbf{S})^{i}\mathbf{r}^{j}-\frac{4}{3}(\mathbf{r}\times\mathbf{S})^{i}\mathbf{v}^{j}+\frac{8}{3}\frac{\delta m}{m}(\mathbf{v}\times\mathbf{\Sigma})^{i}\mathbf{r}^{j}-\frac{4}{3}\frac{\delta m}{m}(\mathbf{r}\times\mathbf{\Sigma})^{i}\mathbf{v}^{j}\biggr\}_{\text{STF}},\\
I_{(\text{2.5PN)}}^{ij} & =\nu\biggl\{\biggl[\biggl(\frac{5}{21}-\frac{5}{7}\nu\biggr)\mathbf{v}\cdot(\mathbf{r}\times\mathbf{S})+\biggl(\frac{5}{21}+\frac{4}{7}\nu\biggr)\frac{\delta m}{m}\mathbf{v}\cdot(\mathbf{r}\times\mathbf{\Sigma})\biggr]\mathbf{v}^{i}\mathbf{v}^{j}\nonumber \\
 & \qquad\qquad+\biggl[\biggl(-\frac{52}{21}+\frac{10}{7}\nu\biggr)\mathbf{v}\cdot(\mathbf{n}\times\mathbf{S})+\biggl(-\frac{62}{21}+\frac{18}{7}\nu\biggr)\frac{\delta m}{m}\mathbf{v}\cdot(\mathbf{n}\times\mathbf{\Sigma})\biggr]\frac{Gm}{r}\mathbf{n}^{i}\mathbf{r}^{j}\nonumber \\
 & \qquad\qquad+\biggl[\biggl(\frac{19}{21}+\frac{167}{21}\nu\biggr)\frac{Gm}{r}+\biggl(-\frac{2}{21}+\frac{2}{7}\nu\biggr)\mathbf{v}^{2}\biggr](\mathbf{v}\times\mathbf{S})^{i}\mathbf{r}^{j}\nonumber \\
 & \qquad\qquad+\biggl[\biggl(-\frac{1}{3}+\frac{20}{3}\nu\biggr)\frac{Gm}{r}+\biggl(-\frac{2}{21}-\frac{20}{7}\nu\biggr)\mathbf{v}^{2}\biggr]\frac{\delta m}{m}(\mathbf{v}\times\mathbf{\Sigma})^{i}\mathbf{r}^{j}\nonumber \\
 & \qquad\qquad+\biggl[\biggl(-\frac{22}{3}-\frac{10}{3}\nu\biggr)\frac{Gm}{r}+\biggl(-\frac{4}{21}+\frac{4}{7}\nu\biggr)\mathbf{v}^{2}\biggr](\mathbf{r}\times\mathbf{S})^{i}\mathbf{v}^{j}\nonumber \\
 & \qquad\qquad+\biggl[\biggl(-\frac{8}{3}-\frac{34}{21}\nu\biggr)\frac{Gm}{r}+\biggl(-\frac{4}{21}+\frac{12}{7}\nu\biggr)\mathbf{v}^{2}\biggr]\frac{\delta m}{m}(\mathbf{r}\times\mathbf{\Sigma})^{i}\mathbf{v}^{j}\nonumber \\
 & \qquad\qquad+\biggl[\biggl(\frac{8}{3}-\frac{16}{3}\nu\biggr)\mathbf{S}\cdot\mathbf{n}+\biggl(\frac{8}{3}-\frac{8}{3}\nu\biggr)\frac{\delta m}{m}\mathbf{\Sigma}\cdot\mathbf{n}\biggr]\frac{Gm}{r}(\mathbf{v}\times\mathbf{n})^{i}\mathbf{r}^{j}\nonumber \\
 & \qquad\qquad+\biggl(\frac{10}{21}-\frac{10}{7}\nu\biggr)r\dot{r}(\mathbf{v}\times\mathbf{S})^{j}\mathbf{v}^{i}+\biggl(\frac{10}{21}-\frac{8}{21}\nu\biggr)\frac{\delta m}{m}r\dot{r}(\mathbf{v}\times\mathbf{\Sigma})^{i}\mathbf{v}^{j}\nonumber \\
 & \qquad\qquad+\biggl(\frac{31}{21}+\frac{19}{21}\nu\biggr)\frac{Gm}{r}\dot{r}(\mathbf{n}\times\mathbf{S})^{j}\mathbf{r}^{i}+\biggl(\frac{5}{3}+\frac{2}{7}\nu\biggr)\frac{Gm}{r}\frac{\delta m}{m}\dot{r}(\mathbf{n}\times\mathbf{\Sigma})^{i}\mathbf{r}^{j}\biggr\}_{\text{STF}}.\label{eq:Iij_NLOSO}
\end{align}
The mass octupole moments are
\begin{align}
I_{\text{(0PN)}}^{ijk} & =-\delta m\nu\{\mathbf{r}^{i}\mathbf{r}^{j}\mathbf{r}^{k}\}_{\text{TF}},\label{eq:mass_oct01}\\
I_{\text{(1PN})}^{ijk} & =-\delta m\nu\biggl\{\Bigl[\Bigl(-\frac{5}{6}+\frac{13\nu}{6}\Bigr)\frac{Gm}{r}+\Bigl(\frac{5}{6}-\frac{19}{6}\nu\Bigr)v^{2}\Bigr]\mathbf{r}^{i}\mathbf{r}^{j}\mathbf{r}^{k}+(-1+2\nu)r\dot{r}\mathbf{r}^{i}\mathbf{r}^{j}\mathbf{v}^{k}+(1-2\nu)r^{2}\mathbf{r}^{i}\mathbf{v}^{j}\mathbf{v}^{k}\biggr\}_{\text{STF}},\\
I_{\text{(1.5PN)}}^{ijk} & =\nu\biggl\{-\frac{9}{2}\frac{\delta m}{m}(\mathbf{v}\times\mathbf{S})^{i}\mathbf{r}^{j}\mathbf{r}^{k}+\Bigl(-\frac{9}{2}+\frac{33}{2}\nu\Bigr)(\mathbf{v}\times\mathbf{\Sigma})^{i}\mathbf{r}^{j}\mathbf{r}^{k}
+3\frac{\delta m}{m}(\mathbf{r}\times\mathbf{S})^{i}\mathbf{r}^{j}\mathbf{v}^{k}+(3-9\nu)(\mathbf{r}\times\mathbf{\Sigma})^{i}\mathbf{r}^{j}\mathbf{v}^{k}\biggr\}_{\text{STF}}.
\end{align}
The current quadrupoles moments are
\begin{align}
J_{\text{(0PN)}}^{ij} & =\nu\delta m\{(\mathbf{v}\times\mathbf{r})^{i}\mathbf{r}^{j}\}_{\text{STF}},\label{eq:curr_quad01}\\
J_{\text{(0.5PN})}^{ij} & =-\frac{3}{2}\nu\{\mathbf{\Sigma}^{i}\mathbf{r}^{j}\}{}_{\text{STF}},\\
J_{\text{(1PN)}}^{ij} & =\nu\delta m\biggl\{\biggl[\biggl(\frac{27}{14}+\frac{15}{7}\nu\biggr)\frac{Gm}{r}+\biggl(\frac{13}{28}-\frac{17}{7}\nu\biggr)\mathbf{v}^{2}\biggr](\mathbf{v}\times\mathbf{r})^{i}\mathbf{r}^{j}+\biggl(\frac{5}{28}-\frac{5}{14}\nu\biggr)r\dot{r}(\mathbf{v}\times\mathbf{r})^{i}\mathbf{v}^{j}\biggr\}_{\text{STF}},\\
J_{\text{(1.5PN)}}^{ij} & =\nu\biggl\{\biggl[\biggl(\frac{61}{28}-\frac{71}{28}\nu\biggr)\frac{Gm}{r}+\biggl(-\frac{2}{7}+\frac{20}{7}\nu\biggr)\mathbf{v}^{2}\biggr]\mathbf{\Sigma}^{i}\mathbf{r}^{j}+\biggl[\frac{10}{7}\frac{Gm}{r}+\frac{13}{28}\mathbf{v}^{2}\biggr]\frac{\delta m}{m}\mathbf{S}^{i}\mathbf{r}^{j}\nonumber \\
 & \qquad+\biggl[-\frac{11}{14}\frac{\delta m}{m}\mathbf{S}\cdot\mathbf{r}+\biggl(-\frac{11}{14}+\frac{47}{14}\nu\biggr)\mathbf{\Sigma}\cdot\mathbf{r}\biggr]\mathbf{v}^{i}\mathbf{v}^{j}+\biggl[\frac{3}{7}\frac{\delta m}{m}\mathbf{S}\cdot\mathbf{v}+\biggl(\frac{3}{7}-\frac{23}{7}\nu\biggr)\mathbf{\Sigma}\cdot\mathbf{v}\biggr]\mathbf{v}^{i}\mathbf{x}^{j}\nonumber \\
 & \qquad+\biggl[-\frac{29}{14}\frac{\delta m}{m}\mathbf{S}\cdot\mathbf{n}+\biggl(-\frac{4}{7}+\frac{31}{14}\nu\biggr)\mathbf{\Sigma}\cdot\mathbf{n}\biggr]\frac{Gm}{r}\mathbf{n}^{i}\mathbf{r}^{j}+\frac{3}{7}\frac{\delta m}{m}r\dot{r}\mathbf{S}^{j}\mathbf{v}^{i}+\biggl(\frac{3}{7}-\frac{16}{7}\nu\biggr)r\dot{r}\mathbf{\Sigma}^{i}\mathbf{v}^{j}\biggr\}_{\text{STF}}.\label{eq:Jij_NLOSO}
\end{align}
The current octupole moments are
\begin{align}
J_{\text{(0PN)}}^{ijk} & =-m\nu(1-3\nu)\{(\mathbf{v}\times\mathbf{r})^{i}\mathbf{r}^{j}\mathbf{r}^{k}\}_{\text{STF}},\\
J_{\text{(0.5PN)}}^{ijk} & =2\nu\biggl\{\mathbf{S}^{i}\mathbf{r}^{j}\mathbf{r}^{k}+\frac{\delta m}{m}\mathbf{\Sigma}^{i}\mathbf{r}^{j}\mathbf{r}^{k}\biggr\}_{\text{STF}}.\label{eq:curr_octo_spin}
\end{align}

\subsection*{NRGR vertices}

The vertices needed to compute the 2.5PN center-of-mass correction
\cite{nrgrs,Porto:2016pyg} are
\begin{align}
S_{H}^{v^{0}} & =-\sum_{A}\frac{m_{A}}{2m_{\text{Pl}}}\int dt_{A}\,H_{00}(x_{A})\label{eq:Hv0},\\
S_{H}^{v^{1}} & =-\sum_{A}\frac{m_{A}}{m_{\text{Pl}}}\int dt_{A}\,v_{A}^{i}H_{0i}(x_{A})\label{eq:Hv1},\\
S_{H\bar{h}_{00}}^{v^{0}} & =\sum_{A}\frac{m_{A}}{4m_{\text{Pl}}^{2}}\int dt_{A}\,H_{00}(x_{A})\bar{h}_{00}(x_{A})\label{eq:Hhv0},\\
S_{H\bar{h}_{00}}^{v^{1}} & =\sum_{A}\frac{m_{A}}{2m_{\text{Pl}}^{2}}\int dt_{A}\,v_{A}^{i}H_{0i}(x_{A})\bar{h}_{00}(x_{A})\label{eq:Hhv1},\\
S_{H}^{Sv^{0}} & =\sum_{A}\frac{1}{2m_{\text{Pl}}}\int dt_{A}\,H_{i0,k}(x_{A})S_{A}^{ik}\label{eq:HSv0},\\
S_{H}^{Sv^{1}} & =\sum_{A}\frac{1}{2m_{\text{Pl}}}\int dt_{A}\,\bigl[H_{ij,k}(x_{A})S_{A}^{ik}v_{A}^{j}+H_{00,k}(x_{A})S_{A}^{0k}\bigr]\label{eq:HSv1},\\
S_{\bar{h}_{00}}^{Sv^{1}} & =\sum_{A}\frac{1}{2m_{\text{Pl}}}\int dt_{A}\,\bar{h}_{00,k}(x_{A})S_{A}^{0k}\label{eq:hSi0v1},\\
S_{H\bar{h}_{00}}^{Sv^{0}} & =\sum_{A}\frac{1}{4m_{\text{Pl}}^{2}}\int dt_{A}\,S_{A}^{ij}H_{j}^{\enskip0}(x_{A})\bar{h}_{00,i}(x_{A})\label{eq:HhSijv0},\\
S_{H\bar{h}_{00}}^{Sv^{1}} & =\sum_{A}\frac{1}{4m_{\text{Pl}}^{2}}\int dt_{A}\,S_{A}^{i0}\bigl[H_{00}(x_{A})\bar{h}_{00,i}(x_{A})+\bar{h}_{00}(x_{A})H_{00,i}(x_{A})+H_{\enskip i}^{l}(x_{A})\bar{h}_{00,l}(x_{A})\bigr].\label{eq:HhSi0v1}
\end{align}
Vertices are expressed using the Minkowski metric.

\bibliographystyle{utphys}
\bibliography{Draft}

\end{document}